\documentclass[11pt,a4paper]{article}
\pdfoutput=1
\usepackage[left=3cm,right=3cm,top=3cm,bottom=3cm]{geometry}

\usepackage[numbers]{natbib}

\usepackage{amssymb,amsmath,mathtools,amsfonts}
\usepackage{mathrsfs}
\usepackage{caption}
\usepackage{float}

\usepackage{color}

\usepackage{bbm}

\usepackage[pdftex]{graphicx} 
\usepackage{subcaption}
\usepackage{booktabs} 

\usepackage{enumitem}

\usepackage{hyperref}

\begin{document}

\title{GARCH-UGH: A bias-reduced approach for dynamic extreme Value-at-Risk estimation in financial time series}

\author{Hibiki Kaibuchi$^{a}$, Yoshinori Kawasaki$^{b}$ \& Gilles Stupfler$^{c}$ }
\date{$^{a}$ {\small The Graduate University for Advanced Studies, 
Japan} \\ $^{b}$ {\small Institute of Statistical Mathematics, The Graduate University of Advanced Studies, Japan} \\ $^{c}$ {\small Univ Rennes, Ensai, CNRS, CREST - UMR 9194, F-35000 Rennes, France}} 
\maketitle

\begin{abstract}

The Value-at-Risk (VaR) is a widely used instrument in financial risk management. The question of estimating the VaR of loss return distributions at extreme levels is an important question in financial applications, both from operational and regulatory perspectives; in particular, the dynamic estimation of extreme VaR given the recent past has received substantial attention. We propose here a two-step bias-reduced estimation methodology called GARCH-UGH (Unbiased Gomes-de Haan), whereby financial returns are first filtered using an AR-GARCH model, and then a bias-reduced estimator of extreme quantiles is applied to the standardized residuals to estimate one-step ahead dynamic extreme VaR. Our results indicate that the GARCH-UGH estimates are more accurate than those obtained by combining conventional AR-GARCH filtering and extreme value estimates from the perspective of in-sample and out-of-sample backtestings of historical daily returns on several financial time series.  \\

 
\end{abstract}

\section{Introduction}
\label{S:1}

A major concern in financial risk management is to quantify the risk associated to high-impact, low-probability extreme losses. The most widely known risk measure is Value-at-Risk (VaR), defined as an extreme quantile of the loss distribution. An estimation of the unconditional VaR (that is, of the common distribution of returns over time, assumed to be stationary) is appropriate for the estimation of potential large levels of loss over the long term, for example with the goal of making long-term investment decisions. On the other hand, the conditional VaR is more appropriate for day-to-day and short-term risk management by capturing the dynamics and the key properties of financial asset returns such as volatility clustering and leptokurtosis. The estimation of the conditional VaR, on which we focus in this paper, therefore gives a better understanding of the riskiness of the portfolio because this riskiness varies with the changing volatility. 

There are two main classes of methods to estimate the conditional VaR. Nonparametric historical simulation relies on observed data directly, and uses the empirical distribution of past losses without assuming any specific distribution, see \cite{danielsson2011}, and \cite{mcneil2015}. Although historical simulation is easy to implement, the estimation of extreme quantiles is difficult as the extrapolation beyond observed returns is impossible because historical simulation essentially assumes that one of the observed returns is expected to be the next period return. By contrast, the parametric approach generally refers to the use of an econometric model of volatility dynamics such as, among many others, the Generalized Autoregressive Conditional heteroskedasticity (GARCH) model of \cite{bollerslev1986}. These models estimate VaRs reflecting the conditional heteroskedasticity of financial data. However, GARCH-type models assuming a parametric distribution of the innovation variable tend to underestimate the risk because this strong parametric assumption is not well-suited to the accurate estimation of heavy-tailedness in the conditional returns of financial time series. 

To overcome the problems of purely nonparametric or parametric estimations of extreme VaR, \cite{mcneil2000} proposes a two-step approach combining a GARCH-type model and Extreme Value Theory (EVT), referred to as GARCH-EVT throughout. EVT focuses only on the tails and allows the extrapolation beyond available data, while traditional econometric models focus on the whole distribution at the expense of consideration of the tails. The key idea of the GARCH-EVT method is to estimate the dynamic extreme VaR by first filtering financial time series with a GARCH-type model to estimate the current volatility, and then by applying the EVT method to the standardized residuals for estimating the tails of the residual distribution. This approach has been widely used to estimate the conditional VaR. For instance \cite{bystrom2004,fernandez2005} find GARCH-EVT to give accurate VaR estimates for standard and extreme quantiles compared with GARCH-type models and unconditional EVT methods on stock market data collected across the US, Latin America, Europe and Asia. Being a two-step procedure based on GARCH-type filtering, the accuracy of the GARCH-EVT approach has been debated. \cite{chavez2005} points out that estimates of the conditional VaR via the GARCH-EVT approach are sensitive to the fitting of a GARCH-type model to the dataset in the first step. On the other hand, \cite{furio2013,jalal2008} have concluded that there is no evidence of any difference in the conditional VaR estimates, regardless of the particular GARCH model selected to filter financial data.

Since the debate on filtering, several modifications of the conventional GARCH-EVT method have been suggested in the literature to provide a more accurate calculation of the residuals before applying the EVT method in the second step. \cite{yi2014} proposes a semiparametric version of GARCH-EVT based on quantile regression. 
\cite{youssef2015} adapts the FIGARCH, HYGARCH and FIAPARCH models to evaluate conditional VaRs for crude oil and gasoline market. \cite{bee2016} proposes an approach called realized EVT where returns are pre-whitened with a high-frequency based volatility model. \cite{zhao2019} develops hybrid time-varying long-memory GARCH-EVT models by using a variety of fractional GARCH models. 
To the best of our knowledge however, little work has been carried out on the EVT step; \cite{ergen2015} uses the skewed $t$-distribution that is fitted to the standardized residuals from the GARCH step in order to recover a fully parametric specification. 
Meanwhile, a purely semiparametric improvement of the EVT step using bias correction methods has not been investigated so far, even though bias correction should be a key concern in extreme value estimation and inference, and has an extensive history (see {\it e.g.} \cite{cai2013}).

This is the contribution of the present paper. More precisely, in the context of the estimation of the one-step ahead dynamic extreme VaR, we develop a new methodology called GARCH-UGH (standing for Unbiased Gomes-de Haan, after \cite{dehaan2016,gomes2002}) which uses an asymptotically unbiased estimator of the extreme quantile applied to the standardized residuals from the GARCH step of \cite{mcneil2000}, instead of the so-called Peaks-Over-Threshold (POT) method in the GARCH-EVT approach. 
We analyze the performance of our approach on four financial time series, which are the Dow Jones, Nasdaq and Nikkei stock indices, and the Japanese Yen/British Pound exchange rate. As we shall illustrate, our results indicate that GARCH-UGH provides substantially more accurate one-step ahead conditional VaRs than the conventional GARCH-EVT and bias-reduced EVT without filtering, based on the performance of the in-sample and out-of-sample backtestings. In addition, our bias-reduction procedure will be designed to be robust to departure from the independence assumption, and as such will be able to handle residual dependence present after filtering in the first step. Our finite-sample results will also illustrate that the GARCH-UGH method leads to one-step ahead conditional VaR estimates that are less sensitive to the choice of sample fraction, and hence mitigates the difficulty in selecting the optimal number of observations for the estimations. Finally, the computational cost of GARCH-UGH is lower than that of conventional GARCH-EVT: the extreme value step in the GARCH-UGH method is semiparametric with an automatic and fast recipe for the estimations of the one-step ahead conditional VaRs, while the GARCH-EVT method is based on fitting a Generalized Pareto Distribution (GPD) to the residuals using Maximum Likelihood Estimation, which may be time-consuming since an appropriate starting value for the estimates has to be found by searching the parameter space.

The rest of the paper is organized as follows. Section~\ref{S:2} presents our framework and methodology. Section~\ref{S:3} first describes the four financial time series used in the empirical analysis, and discusses the performance of our proposed approach through in-sample and out-of-sample backtestings of one-step ahead extreme VaRs compared to existing approaches. Section~\ref{S:Discussion} discusses our findings and perspectives for future work.

\section{The GARCH-UGH method and framework}
\label{S:2}

Let $p_t$ be a daily-recorded price for a stock, index or exchange rate, and let $X_t=-\log(p_t/p_{t-1})$ be the negative daily log-return on this price. We assume that the dynamics of $X_t$ are governed by 
\begin{equation} 
\label{E:2.1.1}
X_t = \mu_{t} + \sigma_{t}Z_t,
\end{equation}
where $\mu_t\in \mathbb{R}$ and $\sigma_t>0$ denote the (conditional) mean and standard deviation, and the innovations $Z_t$ form a strictly stationary white noise process, that is, they are i.i.d.~with zero mean, unit variance and common marginal distribution function $F_Z$. We assume that for each $t$, $\mu_t$ and $\sigma_t$ are measurable with respect to the $\sigma-$algebra $\mathscr{F}_{t-1}$ representing the information about the return process available up to time $t-1$. 

We are concerned with estimating extreme conditional quantiles of these negative log-returns. Recall that for a probability level $\tau \in (0,1)$, the $\tau$th unconditional quantile of a distribution $F$ is 
$
q_{\tau} = \inf \{x \in \mathbb{R} : F(x) \geq \tau \}.
$
Here we focus on the one-step ahead quantile, that is, the estimation of the conditional extreme quantile of $X_{t+1}$ given $\mathscr{F}_t$. In this case, by location equivariance and positive homogeneity of quantiles, the one-step ahead conditional extreme quantile (or VaR) of $X_{t+1}$ can be written as 
\begin{equation} 
\label{E:2.1.2}
q_{\tau}(X_{t+1} \mid \mathscr{F}_t) = \mu_{t+1} + \sigma_{t+1} q_{\tau}(Z),
\end{equation}
where $q_{\tau}(Z)$ is the common $\tau$th quantile of the marginal distribution of the innovations $Z_t$. The problem of estimating $q_{\tau}(X_{t+1} \mid \mathscr{F}_t)$ can then be tackled by estimating the mean and standard deviation components $\mu_{t+1}$ and $\sigma_{t+1}$ and the unconditional quantile $q_{\tau}(Z)$. Given estimates $\hat{\mu}_{t+1}$, $\hat{\sigma}_{t+1}$ and $\hat{q}_{\tau}(Z)$ of these quantities, an estimate of $q_{\tau}(X_{t+1} \mid \mathscr{F}_t)$ is then 
\[
\hat{q}_{\tau}(X_{t+1} \mid \mathscr{F}_t) = \hat{\mu}_{t+1} + \hat{\sigma}_{t+1} \hat{q}_{\tau}(Z).
\]
In calculating this estimate, there are three main difficulties. First, one has to estimate $\mu_{t+1}$ and $\sigma_{t+1}$, which supposes that an appropriate model and estimation method have to be chosen. Second, the innovations $Z_t$ are unobserved, which means that the estimation of $q_{\tau}(Z)$ has to be based on residuals following the estimation of $\mu_{t+1}$ and $\sigma_{t+1}$. A third difficulty is specific to our context: we wish here to estimate a dynamic extreme VaR, that is, a conditional quantile $q_{\tau}(X_{t+1} \mid \mathscr{F}_t)$ with $\tau$ very close to 1. In such contexts, it is well-known that traditional nonparametric estimators become inconsistent (see for example the monographs by \cite{beirlant2004,embrechts1997}), and adapted extrapolation methodologies have to be employed. 

Our GARCH-UGH method combines estimation of the mean and standard deviation in a GARCH-type model with a flexible bias-reduced extrapolation methodology for the estimation of $q_{\tau}(Z)$ using the residuals obtained after estimation of the model structure. We describe these two steps successively below. 

\subsection{GARCH step}
\label{S:GARCHstep}

In order to estimate $\mu_{t+1}$ and $\sigma_{t+1}$, one should select a particular model in the class (\ref{E:2.1.1}). Many different models for volatility dynamics have been used in the literature of GARCH-EVT approach, as we highlighted in our literature review in Section~\ref{S:1}. Here we use an AR(1) model for the dynamics of the conditional mean, and a parsimonious but effective GARCH(1,1) model for the volatility, as in the original GARCH-EVT approach; this will allow us to subsequently illustrate how improving the second, EVT-based step can result in more accurate estimates.

We thus model the conditional mean of the series by $\mu_t = \phi X_{t-1}$, for some $\phi\in (-1,1)$, and the conditional variance of the mean-adjusted series $\epsilon_{t} = X_{t}- \mu_{t}$ by 
$
\sigma^{2}_{t}= \kappa_{0} + \kappa_{1}\epsilon^{2}_{t-1} + \kappa_{2} \sigma^{2}_{t-1}, 
$
where $\kappa_{0}, \kappa_{1}, \kappa_{2} >0$. Necessary and sufficient conditions for the stationarity of a model following GARCH(1,1) dynamics are given in Chapter~2 of~\cite{francq2010}; the condition $\kappa_{1} + \kappa_{2} < 1$ is a simple sufficient condition guaranteeing stationarity. The model is therefore the AR(1)-GARCH(1,1) model
\begin{equation}
\label{E:model}
X_t = \mu_{t} + \sigma_{t}Z_t, \ \mbox{ with } \mu_t = \phi X_{t-1} \mbox{ and } \sigma^{2}_{t}= \kappa_{0} + \kappa_{1} (X_{t-1}- \mu_{t-1})^{2} + \kappa_{2} \sigma^{2}_{t-1}. 
\end{equation}  
In Equation~\eqref{E:model}, the innovations $Z_t$ are i.i.d.~with zero mean, unit variance. 

In order to make one-step ahead predictions at time $t$, we fix a memory $n$ so that at the end of time $t$, the financial data consist of the last $n$ negative log-returns $X_{t-j}$, for $0\leq j\leq n-1$. We then fit the AR(1)-GARCH(1,1) model to the data $(X_{t-n+1}, \ldots, X_{t-1}, X_t)$ using Gaussian Quasi-Maximum Likelihood Estimation (QMLE), that is, by maximizing the likelihood constructed by assuming that the innovations $Z_t$ are i.i.d.~Gaussian with zero mean and unit variance. While of course the innovations $Z_t$ will not be Gaussian in general (and indeed in our UGH step we shall assume that they are heavy-tailed), the QMLE method yields a consistent and asymptotically normal estimator, see for example~\cite{francq2004} for a theoretical analysis. One may also put a strong heavy-tailed parametric specification on $Z_t$, such as assuming that they are location-scale Student distributed; this was tried in our analysis of financial log-returns but did not improve results substantially. 

Let $(\hat{\phi}, \hat{\kappa}_0, \hat{\kappa}_1, \hat{\kappa}_2)$ be the Gaussian QMLE estimates. Choosing sensible starting values for $\hat{\epsilon}^{2}_{t-n}$ and $\hat{\sigma}^{2}_{t-n}$ (for example, constant values as in Section~7.1 of~\cite{francq2010}), estimates of the conditional mean and the conditional standard deviation, $(\hat{\mu}_{t-n+1}, \ldots, \hat{\mu}_{t-1}, \hat{\mu}_t)$ and $(\hat{\sigma}_{t-n+1}, \ldots, \hat{\sigma}_{t-1}, \hat{\sigma}_t)$ respectively, can be calculated from Equation~\eqref{E:model} recursively. This leads to the residuals
\[
(\hat{Z}_{t-n+1}, \ldots,\hat{Z}_{t}) = \bigg(\dfrac{X_{t-n+1} - \hat{\mu}_{t-n+1}}{\hat{\sigma}_{t-n+1}}, \ldots, \dfrac{X_{t} - \hat{\mu}_{t}}{\hat{\sigma}_{t}} \bigg).
\]
We end this step by calculating the estimates of the conditional mean and standard deviation for time $t+1$, which are the obvious one-step ahead forecasts, as follows:
\begin{align*}
 \hat{\mu}_{t+1}  &= \hat{\phi} X_{t}, \\
 \hat{\sigma}_{t+1}  &= \sqrt{\hat{\kappa}_{0}+ \hat{\kappa}_{1}\hat{\epsilon}^{2}_{t} + \hat{\kappa}_{2}\hat{\sigma}^{2}_{t}},
\end{align*}
where $\hat{\epsilon}_{t} = X_{t} - \hat{\mu}_{t}$. In summary, this first GARCH step of the method consists in fitting an AR(1)-GARCH(1,1) model to the negative log-returns at a certain past time horizon $n$ (not too small so that the method produces reasonable results, and not too large so that the AR-GARCH model is believable over this time period), using a Gaussian QMLE, leading to forecasts $\hat{\mu}_{t+1}$ and $\hat{\sigma}_{t+1}$ and standardized residuals $\hat{Z}_{t-j}$, $0\leq j\leq n-1$. 

\subsection{UGH step}
\label{S:UGHstep}

With standardized residuals at our disposal, we can now discuss the estimation of $q_{\tau}(Z)$. The residuals $\hat{Z}_{t-j}$, $0\leq j\leq n-1$, approximate the true unobservable $Z_{t-j}$. Assume that the underlying distribution of these $Z_{t-j}$ is heavy-tailed, that is  
\begin{equation}
\label{E:2.3.1.1}
\lim_{t \rightarrow \infty} \dfrac{U(tz)}{U(t)} = z^{\gamma}, \ \forall z > 0, \ \mbox{ where } U(t)=q_{1-t^{-1}}(Z). 
\end{equation}
In other words, we assume the tail of the innovations to be approximately Pareto, with the index $\gamma$ tuning how heavy the tail is. This assumption is ubiquitous in actuarial and financial risk management (see e.g.~p.9 of~\cite{embrechts1997} and p.1 of~\cite{resnick2007}). It makes it possible to construct extrapolated extreme quantile estimators: the classical Weissman quantile estimator (see~\cite{weissman1978}) of a quantile $q_{\tau}(Z) = q_{1-p}(Z)$ with $p=1-\tau$ close to 0 is then 
\begin{equation}
\label{E:Weissman}
\overline{q}_{1-p}(Z) = \bigg(\dfrac{k}{np} \bigg)^{\overline{\gamma}_k} Z_{n-k,n} 
\end{equation}
where $Z_{1,n}\leq Z_{2,n}\leq \cdots \leq Z_{n,n}$ are the order statistics from $Z_{t-n+1},\ldots,Z_t$ and $\overline{\gamma}_k$ is a consistent estimator of $\gamma$. The tuning parameter $k$ denotes the effective sample size for the estimation: this parameter should be chosen not too small, so that the variance of the estimator is reasonable, but also not too large so that the bias coming from the use of the extrapolation relationship~\eqref{E:2.3.1.1} does not dominate. The most common estimator $\overline{\gamma}_k$ of $\gamma$ is the Hill estimator (introduced in~\cite{hill1975}): 
\begin{equation}
\label{E:Hill}
\overline{\gamma}_k = \overline{\gamma}^{H}_k = \dfrac{1}{k} \sum_{i=1}^{k} \log Z_{n-i+1,n} - \log Z_{n-k,n}. 
\end{equation}
The Hill and Weissman estimators can be shown to be asymptotically Gaussian (see for example Chapters 3 and 4 in~\cite{dehaan2006}). A reasonable idea to define an estimator of $q_{1-p}(Z)$ in our context is then to use the estimators defined in Equations~\eqref{E:Weissman} and~\eqref{E:Hill} with the order statistics of the residuals, $\hat{Z}_{n-j,n}$, in place of the unobservable $Z_{n-j,n}$. 

The choice of the parameter $k$ requires solving a bias-variance tradeoff for which there is no straightforward approach. Indeed, with a low $k$, the estimators use observations that are very informative about the extremes, but their low number results in a high variance. With a high $k$, the variance is reduced, but at the cost of taking into account observations that are further into the bulk of the distribution and thus carry bias. One possible way to make the choice of $k$ easier is to work on correcting this bias. This can be done under the following so-called second-order condition on $U$: 
\begin{equation}
\label{E:2ndorder}
\lim_{t \rightarrow \infty} \frac{1}{A(t)} \left( \dfrac{U(tz)}{U(t)} -z^{\gamma} \right) = z^{\gamma} \dfrac{z^{\rho}-1}{\rho}, \ \forall z > 0, 
\end{equation} 
where $\rho\leq 0$ is called the second-order parameter and $A$ is a positive or negative function converging to 0 at infinity, such that $|A|$ is regularly varying with index $\rho$. The function $A$ therefore controls the rate of convergence in Equation~\eqref{E:2.3.1.1}: the larger $|\rho|$ is, the faster $|A|$ converges to~0, and the smaller the error in the approximation of the right tail of $U$ by a Pareto tail is. This makes it possible to precisely quantify the bias of the Hill and Weissman estimators, and to correct for this bias by estimating the function $A$ and the parameter $\rho$. This results in bias-corrected Hill and Weissman estimators for which the selection of $k$ is typically much easier because their performance is much more stable. 

Our idea in this second, UGH step is to apply such bias-corrected estimators constructed in~\cite{dehaan2016} (and built on second-order parameter estimators of~\cite{gomes2002}, hence the name UGH, for Unbiased Gomes-de Haan) to our residuals obtained from the GARCH step. Let $m$ be the number of positive residuals and, for any positive number $\alpha\notin\{ 1/2,1 \}$, set
\begin{equation*}
\begin{split}
M_{k}^{(\alpha)} & = \dfrac{1}{k} \sum_{i=1}^{k} (\log \hat{Z}_{n-i+1,n} - \log \hat{Z}_{n-k,n})^{\alpha}, \\
R_{k}^{(\alpha)} & = \dfrac{M_{k}^{(\alpha)} - \Gamma (\alpha +1)(M_{k}^{(1)})^{\alpha}}{M_{k}^{(2)} - 2(M_{k}^{(1)})^2}, \\
S_{k}^{(\alpha)} & = \dfrac{\alpha (\alpha +1)^2 \Gamma^2 (\alpha)}{4 \, \Gamma (2 \alpha)} \dfrac{R_{k}^{(2\alpha)}}{(R_{k}^{(\alpha +1)})^2}, \\
s^{(\alpha)}(\rho) & = \dfrac{\rho^2 (1- (1-\rho)^{2 \alpha} - 2 \alpha \rho (1 - \rho)^{2 \alpha -1}  )}{(1- (1 - \rho)^{\alpha +1} - (\alpha +1) \rho (1 - \rho)^{\alpha}   )^2}.
\end{split}
\end{equation*}
Our estimator of $\rho$ motivated by~\cite{gomes2002} is  
\[
\hat{\rho}_{k}^{(\alpha)} = (s^{(\alpha)})^\leftarrow (S_{k}^{(\alpha)}),
\]
where $\alpha$ is a positive tuning parameter and ${}^\leftarrow$ denotes the generalized (left-continuous) inverse. The version of $\hat{\rho}_{k}^{(\alpha)}$ with the true innovations $Z_k$ instead of the residuals is known to be consistent under a so-called third-order condition which further strengthens~\eqref{E:2ndorder}. Here we choose $\alpha =2$ since, on fully observed data, this appears to yield the smallest mean squared error following the numerical simulations of \cite{gomes2002}. This results in the estimator 
%
\[
\hat{\rho}_{k}^{(2)} = \dfrac{-4 + 6S_{k}^{(2)} + \sqrt{3S_{k}^{(2)}-2}}{4S_{k}^{(2)}-3}
\]
%
provided $2/3 \leq S_{k}^{(2)} \leq 3/4$, where 
\[
S_{k}^{(2)} = \dfrac{3}{4} \, \dfrac{[M_{k}^{(4)}- 24(M_{k}^{(1)})^{4} ] [M_{k}^{(2)} - 2(M_{k}^{(1)})^{2} ] }{[M_{k}^{(3)}- 6(M_{k}^{(1)})^{3} ]^{2}}.
\]
[This expression corrects a typo in p.375 of~\cite{dehaan2016}.] It is clear that $\hat{\rho}_{k}^{(2)}$ does not exist if $S_{k}^{(2)} \notin [ 2/3, 3/4]$. In practice, we select the value of $k$ in this estimator by setting
\begin{equation}
\label{E:krho}
k_\rho = \sup \bigg\{k : k \leq \min \bigg(m-1, \dfrac{2m}{\log \log m}  \bigg) \ \text{and} \ \hat{\rho}_k^{(2)} \ \text{exists} \bigg\}.
\end{equation}
Here $m$ is the number of positive observations in the sample. The intuition is that even though this estimator of $\rho$ requires a choice of $k$, this choice should be different from its counterpart used in the estimation of $\gamma$, and indeed intuitively the value of $k$ in the estimator of $\rho$ should be rather high in order to allow the methodology to identify the bias coming from including observations belonging to the bulk of the distribution (which correspond to a high $k$). We then estimate $\rho$ by $\hat{\rho}_{k_{\rho}}=\hat{\rho}_{k_{\rho}}^{(2)}$. If the set on the right-hand side of~\eqref{E:krho} is empty, we define $\hat{\rho}_{k_{\rho}}=-1$ as recommended in p.117 of Section 4.5.1 in~\cite{beirlant2004}. The parameter $\gamma$ is then estimated using the residual-based version of the bias-corrected Hill estimator introduced in~\cite{dehaan2016}: 
%
\[
\hat{\gamma}_{k,k_\rho} = \hat{\gamma}_{k}^{H} - \dfrac{ M_{k}^{(2)} - 2(\hat{\gamma}_{k}^{H})^{2}}{2 \hat{\gamma}_{k}^{H} \hat{\rho}_{k_\rho} (1- \hat{\rho}_{k_\rho})^{-1} }. 
\]
%
Our residual-based, bias-corrected estimator of unconditional extreme quantiles is then given by
\begin{equation}
\label{E:2.3.2.3}
\hat{q}_{1-p}(Z) =  \hat{Z}_{n-k,n} \bigg(\dfrac{k}{np} \bigg)^{\hat{\gamma}_{k, k_\rho}} \times \Bigg(1 - \dfrac{\big[M_k^{(2)}- 2(\hat{\gamma}_k^{H})^{2}\big] [1- \hat{\rho}_{k_\rho}]^{2}}{2 \hat{\gamma}_{k}^{H} \hat{\rho}_{k_\rho}^{2}}  \Bigg[1- \bigg(\dfrac{k}{np}  \bigg)^{\hat{\rho}_{k_\rho}}  \Bigg]\Bigg).
\end{equation}
This corresponds to a slightly different version of the estimator in Section~4.3 of~\cite{dehaan2016}, given later by~\cite{chavez2018}, who pointed out a mistake in the analysis of~\cite{dehaan2016}. The versions of these estimators for fully observed data work when this data is weakly serially dependent, as shown in~\cite{chavez2018}. As such, our proposed method will be robust to the presence of residual dependence after filtering and to model misspecification in the sense of~\cite{hillJB2015}. We shall also show that the choice of $k$ for this estimator is not as crucial in finite samples as for the traditional Hill estimator, because this estimator has reasonably good performance across a large range of values of $k$. 

\subsection{Summary and output of the GARCH-UGH method}
\label{S:summary}

The GARCH-UGH approach may be summarized by the following two successive steps:

\begin{enumerate}
\item GARCH step: based on $n$ previous observations at time $t$, fit an AR(1)-GARCH(1,1) model to the negative daily log-returns data using a Gaussian QMLE. Obtain $\hat{\mu}_{t+1}$ and $\hat{\sigma}_{t+1}$ using the fitted model and compute standardized residuals.
\item UGH step: substitute the standardized residuals into the asymptotically unbiased tail quantile estimator $\overline{q}_{\tau}(Z)$, resulting in the estimator $\hat{q}_{1-p}(Z)$ in~\eqref{E:2.3.2.3}.
\end{enumerate}

Combining the two steps results in the final GARCH-UGH estimator
\[
\hat{q}_{\tau}(X_{t+1} \mid \mathscr{F}_t) = \hat{\mu}_{t+1} + \hat{\sigma}_{t+1} \hat{q}_{\tau}(Z).
\]
The goal of our real data analysis is to examine the finite-sample performance of this estimator for low exceedance probabilities, that is, $\tau$ close to 1. 

\section{Empirical analysis of four financial time series}
\label{S:3}

\begin{figure}

\begin{subfigure}{0.55\textwidth}
\includegraphics[width=0.9\linewidth, height=6cm]{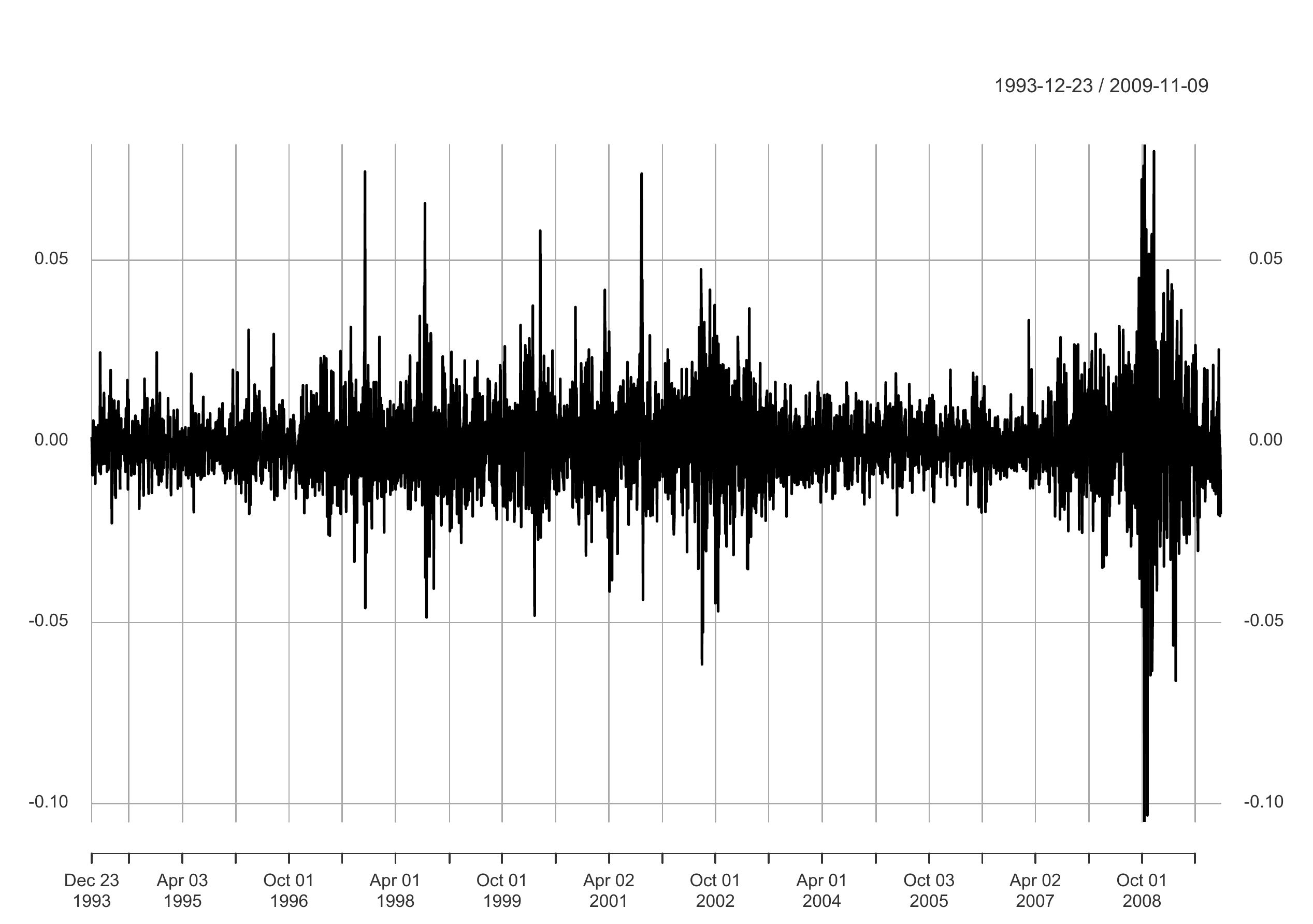} 
\caption{DJ (23/12/1993 to 09/11/2009) 
}
\label{F:1.1}
\end{subfigure}
\begin{subfigure}{0.55\textwidth}
\includegraphics[width=0.9\linewidth, height=6cm]{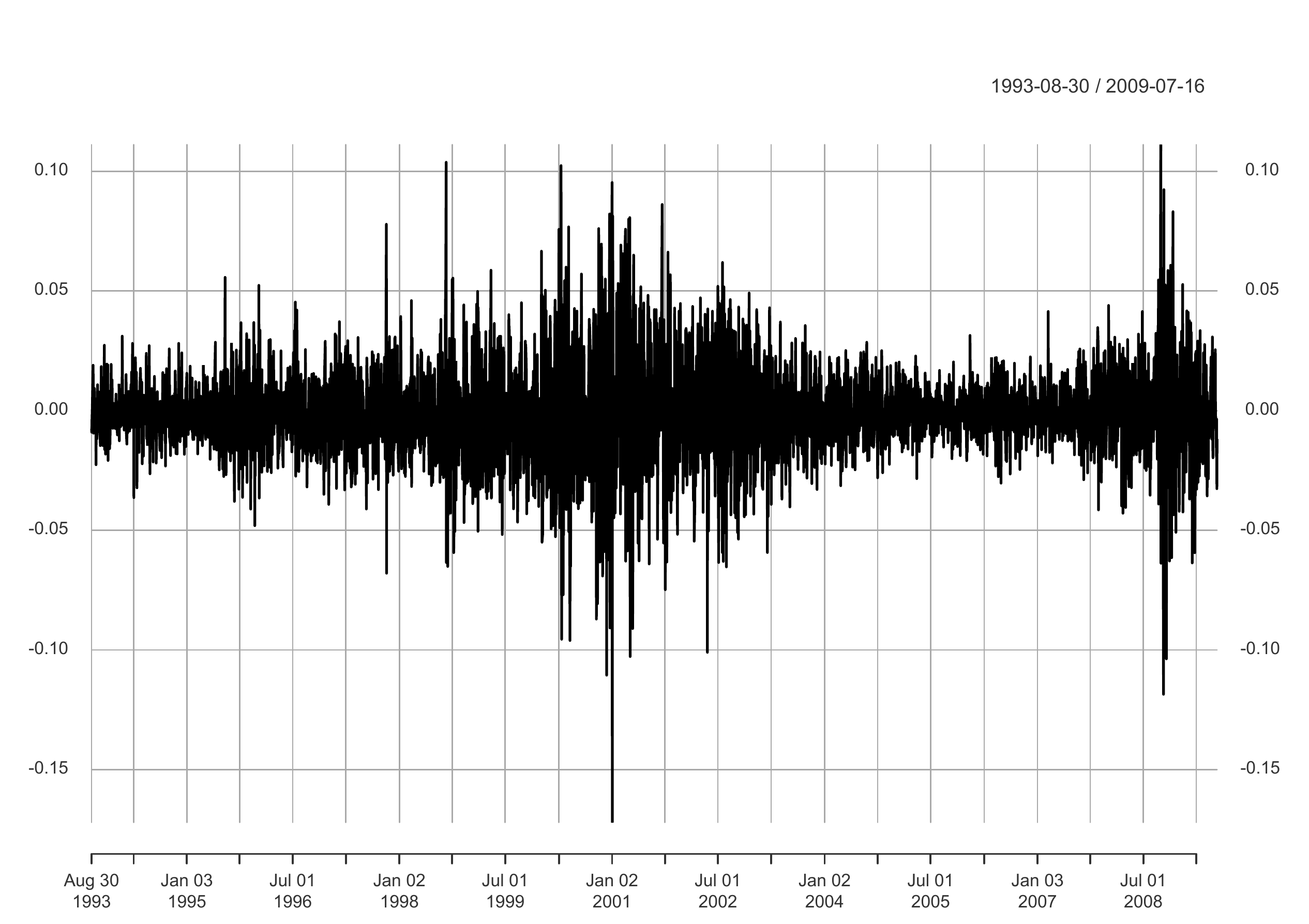}
\caption{NASDAQ (30/08/1993 to 16/07/2009) 
}
\label{F:1.2}
\end{subfigure}
\begin{subfigure}{0.55\textwidth}
\includegraphics[width=0.9\linewidth, height=6cm]{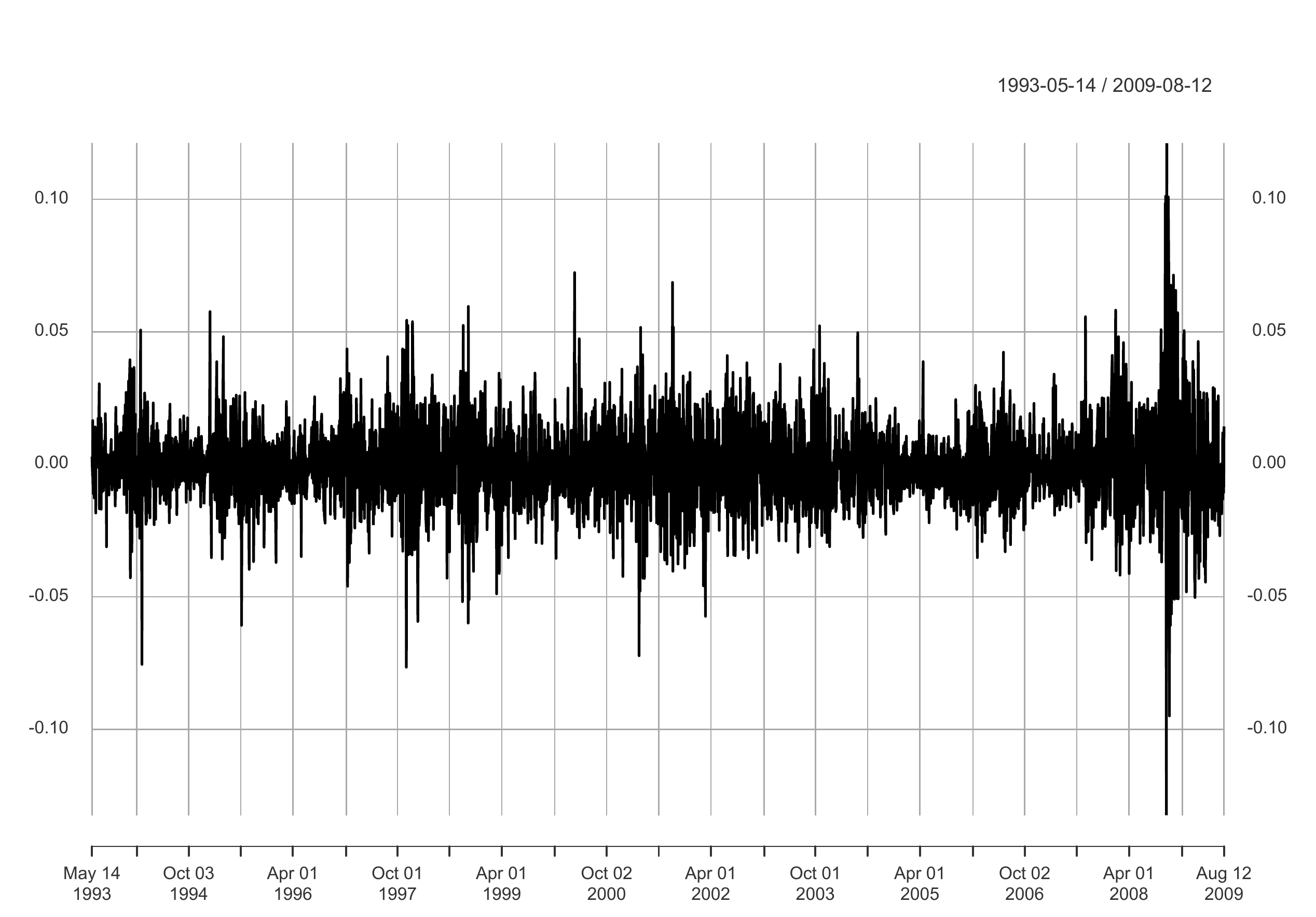} 
\caption{NIKKEI (14/05/1993 to 12/08/2009) 
}
\label{F:1.3}
\end{subfigure}
\begin{subfigure}{0.55\textwidth}
\includegraphics[width=0.9\linewidth, height=6cm]{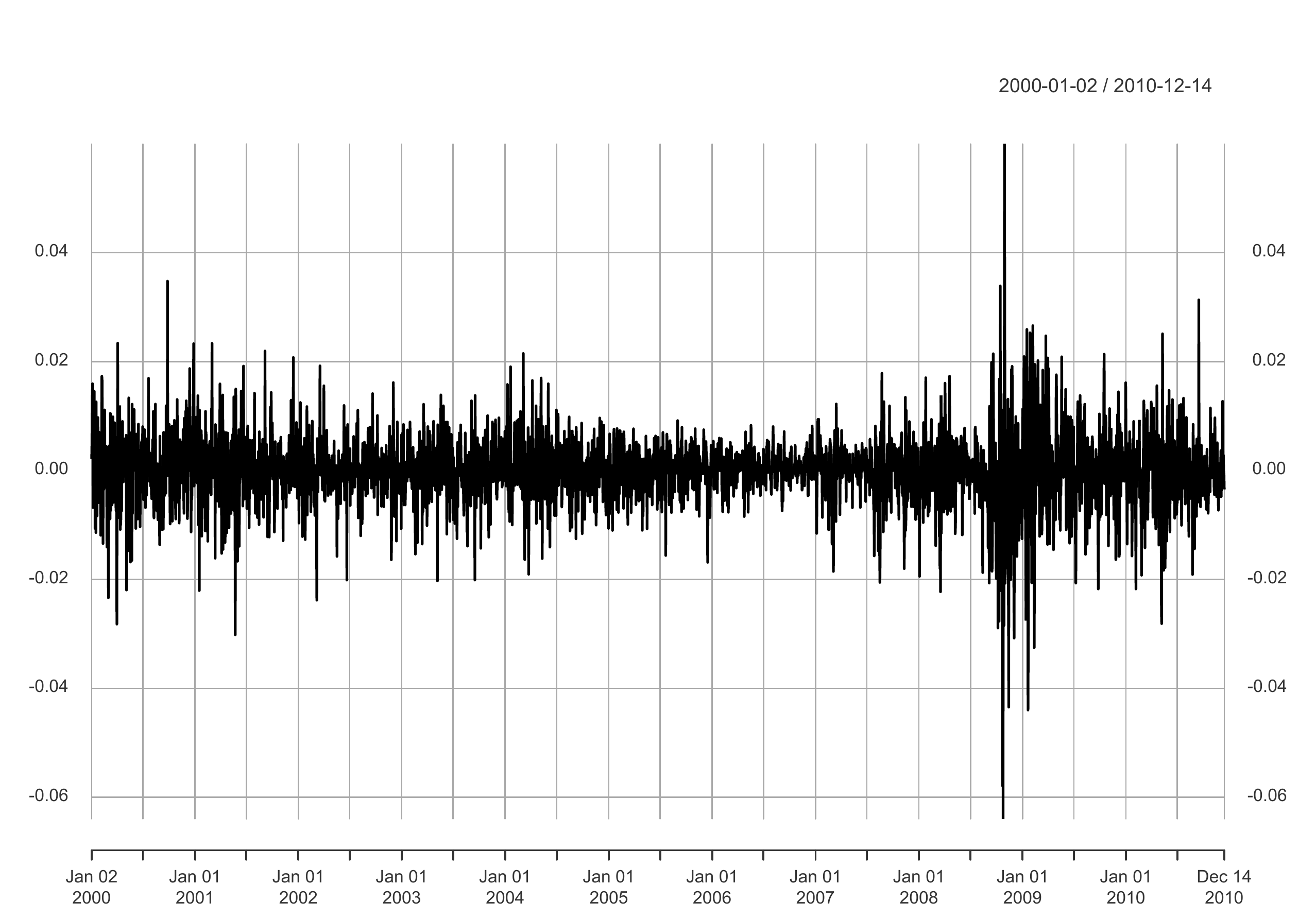}
\caption{JPY/GBP (02/01/2000 to 14/12/2010) 
}
\label{F:1.4}
\end{subfigure}

\caption{Daily negative log-returns of four financial time series: DJ, NASDAQ, NIKKEI and JPY/GBP.}
\label{F:1}
\end{figure}

We consider historical daily negative log-returns of three financial indices and an exchange rate, all made of $n=4000$ observations: 
\begin{itemize}
\item The Dow Jones Industrial Average (DJ) from 23 December 1993 to 9 November 2009; 
\item The Nasdaq Stock Market Index (NASDAQ) from 30 August 1993 to 16 July 2009; 
\item The Nikkei $225$ (NIKKEI) from 14 May 1993 to 12 August 2009; 
\item The Japanese Yen-British Pound exchange rate (JPY/GBP) from 2 January 2000 to 14 December 2010. 
\end{itemize}
The data have been taken from the R package {\tt qrmdata}~\cite{qrmdata} and are represented in Figure~\ref{F:1}. The graphs show that these negative log-returns are extremely volatile around the $2007$-$2008$ financial crisis, which created a succession of extreme positive and negative returns over short time horizons. A noticeable degree of volatility clustering is also detected from a visual inspection of Figure~\ref{F:1}, revealing the presence of the heteroskedasticity. 

Descriptive statistics and basic statistical tests applied to the negative log-returns on the four financial time series are reported in Table~\ref{T:1}. According to the descriptive statistics, the mean of the negative log-returns of all series are close to zero, 
and negative log-returns are 
leptokurtic. The Jarque-Bera test statistics indicate that the Gaussian distribution is not suitable for any of these series of negative log-returns. All four series pass the augmented Dickey-Fuller (ADF) test, indicating that they can be considered stationary for modeling purposes. The Ljung-Box test applied to the squared negative log-returns, with orders $1$ and $10$, rejects the null hypothesis of no autocorrelation, indicating the presence of substantial conditional heteroskedasticity in all series. This provides justification for our use of GARCH-type models with these data. 
 
We compare the performance of our GARCH-UGH method with two benchmarks: 
\begin{itemize} 
\item The bias-reduced UGH method without filtering: this method applies the UGH step directly to the series $X_t$. 
\item The conventional GARCH-EVT method as described in~\cite{mcneil2000}. This consists, first, in the same filtering step as described in Section~\ref{S:GARCHstep}. Standardized residuals are then recorded and a Generalized Pareto distribution is fitted using a maximum likelihood estimator, thus producing a VaR estimate $\tilde{q}_{\tau}(Z)$. This method therefore differs from ours as far as the extreme value step is concerned. 
\end{itemize}
A comparison with the UGH method (without filtering) allows us to see how effective filtering is, and a comparison to the GARCH-EVT method (not featuring bias reduction) will illustrate the benefit of bias reduction at the extreme value step after filtering. In Section~\ref{S:4.1}, we explain how we carry out backtesting of the performance of one-step ahead conditional VaR estimators provided by each approach. This is then followed by in-sample and out-of-sample evaluations of one-step ahead conditional VaR estimates at different $\tau$ levels and choices of $k$ in Sections~\ref{S:4.2} and~\ref{S:4.3}, respectively: in-sample estimation investigates the fit of the approaches to high volatile returns, while out-of-sample estimation tests how well the method predicts extreme VaR.

\subsection{Statistical framework for VaR backtesting}
\label{S:4.1}

Backtesting is carried out to examine the accuracy of the one-step ahead conditional VaR estimates. It compares the ex-ante VaR estimates $\hat{q}_{\tau}(X_t \mid \mathscr{F}_{t-1})$ with the ex-post realized negative log-returns in a time window $W_T$, with a VaR violation at time $t$ said to occur whenever $x_t > \hat{q}_{\tau}(X_t \mid \mathscr{F}_{t-1})$. Define a hit sequence of VaR violations as $I_t=\mathbbm{1}\{ x_t > \hat{q}_{\tau}(X_t \mid \mathscr{F}_{t-1}) \}$. If a VaR estimation method is accurate, then the sequence $(I_t)$ should approximately be an independent sequence of Bernoulli variables with success probability $p=1-\tau$. 
Both the distributional and independence properties are equally important. A VaR estimation method with too few VaR violations will tend to overestimate risk and therefore to be excessively conservative in financial terms, while too many VaR violations mean that risk is underestimated, leading to insufficient provision of capital and therefore potential insolvency in case of large losses. Besides, a violation of the independence property typically arises when there is a clustering of VaR violations, which indicates a model that does not represent volatility clustering well enough.

In order to test the distributional assumption, we use the unconditional coverage test proposed by \cite{kupiec1995} (also known as Kupiec test or POF test, for Proportion Of Failures): fix a time window $W_T$, let $N = \sum_{t\in W_T} I_t$ be the observed number of VaR violations over $W_T$ and $p$ be the theoretical violation rate. The Kupiec test statistic is the likelihood ratio (LR) statistic given by 
\[
\text{LR}_{\text{uc}} = -2 \log \{p^N (1-p)^{T-N} \} + 2 \log \bigg\{\bigg(\dfrac{N}{T}\bigg)^N  \bigg(1- \dfrac{N}{T}\bigg)^{T-N}\bigg\}.
\]
Under the null hypothesis that the $I_t$ are independent and Bernoulli distributed with success probability $p$, the test statistic $\text{LR}_{\text{uc}}$ is asymptotically $\chi^{2}$ distributed with 1 degree of freedom. The Kupiec test rejects this null hypothesis with asymptotic type I error $\alpha$ when $\text{LR}_{\text{uc}}>\chi_{1,1-\alpha}^2$, where $\chi_{1,1-\alpha}^2$ is the $(1-\alpha)-$quantile of the $\chi^2$ distribution with 1 degree of freedom. 

One way to test the independence property is to use another likelihood ratio test called the conditional coverage test, proposed by Christoffersen in~\cite{christoffersen1998}. This test is based on testing for first-order Markov dependence, with the test statistic being given by 
\[
\text{LR}_{\text{cc}} = -2 \log \{p^N (1-p)^{T-N} \} + 2 \log \{ \hat{\pi}_{00}^{N_{00}} \hat{\pi}_{01}^{N_{01}} \hat{\pi}_{10}^{N_{10}} \hat{\pi}_{11}^{N_{11}} \}.
\]
Here $N_{ij}=\sum_{t\in W_T} \mathbbm{1}\{ I_{t+1}=j, I_t=i \}$ and $\hat{\pi}_{ij} = N_{ij}/(N_{i0}+N_{i1})$. Under the null hypothesis that the sequence $(I_t)$ is independent and identically distributed as Bernoulli with parameter $p$, the test statistic $\text{LR}_{\text{cc}}$ is asymptotically $\chi^{2}$ distributed with 2 degrees of freedom, and the conditional coverage test then rejects this null hypothesis with asymptotic type I error $\alpha$ when $\text{LR}_{\text{cc}}>\chi_{2,1-\alpha}^2$ (the $(1-\alpha)-$quantile of the $\chi^2$ distribution with 2 degrees of freedom). Strictly speaking the conditional coverage test only assesses departure from either independence or stationarity, but in fact 
\begin{align*}
\text{LR}_{\text{cc}} &= \text{LR}_{\text{uc}} + \text{LR}_{\text{ind}} \\
\mbox{with } \text{LR}_{\text{ind}} &= -2 \log \bigg\{\bigg(\dfrac{N}{T}\bigg)^N  \bigg(1- \dfrac{N}{T}\bigg)^{T-N}\bigg\} + 2 \log \{ \hat{\pi}_{00}^{N_{00}} \hat{\pi}_{01}^{N_{01}} \hat{\pi}_{10}^{N_{10}} \hat{\pi}_{11}^{N_{11}} \}.
\end{align*}
The quantity $\text{LR}_{\text{ind}}$ is nothing but a likelihood ratio test statistic of independence versus nontrivial first-order Markov dynamics of the sequence $(I_t)$, which rejects independence of $(I_t)$ provided $\text{LR}_{\text{uc}}>\chi_{1,1-\alpha}^2$. Since $\chi_{2,1-\alpha}^2 = \chi_{1,1-\alpha}^2 + \chi_{1,1-\alpha}^2$, checking stationarity and independence via the pair of test statistics $(\text{LR}_{\text{uc}},\text{LR}_{\text{ind}})$ is exactly equivalent to checking them via the unconditional and conditional coverage tests. We therefore use below both the unconditional and conditional coverage tests to assess the performance of our dynamic extreme VaR estimators. This constitutes a backtesting approach in the spirit of the one suggested by the Basel Committee on Banking Supervision.

\subsection{In-sample dynamic extreme VaR estimation and backtesting}
\label{S:4.2}

We start by estimating in-sample one-step ahead conditional VaRs $q_{\tau}(X_{t+1} \mid \mathscr{F}_t)$ for $\tau \in \{0.99, 0.995, 0.999\}$. For these in-sample evaluations, all methods (GARCH-UGH, UGH without filtering, and GARCH-EVT without bias reduction) are implemented on a fixed in-sample testing window $W_T$, which consists of $3000$ observations; this follows advice by \cite{danielsson2011} which suggests that this testing window $W_T$ should cover at least $4$ years of data, or approximately $1000$ observations, for a reliable statistical analysis. Specifically, we use: 
\begin{itemize}
\item The time period from 8 December 1997 to 9 November 2009 for the Dow Jones,  
\item The time period from 13 August 1997 to 16 July 2009 for the Nasdaq,
\item The time period from 29 May 1997 to 12 August 2009 for the Nikkei,
\item The time period from 28 September 2002 to 14 December 2010 for the JPY/GBP exchange rate.
\end{itemize}
This allows us to focus on extreme VaR estimation around the $2007$-$2008$ financial crisis, of which a consequence was a succession of extremely large negative log-returns in a very short timeframe. This should be considered a challenging problem.

In each case, we implement the three methods on these $3000$ observations. The GARCH-UGH and GARCH-EVT methods filter the data using an AR(1)-GARCH(1,1) model $X_t = \mu_{t} + \sigma_{t}Z_t$, then estimate $q_{\tau}(Z)$ on the basis of the residuals obtained from this filtering, before obtaining the final extreme VaR estimate as $\hat{q}_{\tau}(X_{t+1} \mid \mathscr{F}_t) = \hat{\mu}_{t+1} + \hat{\sigma}_{t+1} \hat{q}_{\tau}(Z)$ (for the GARCH-UGH method) and $\tilde{q}_{\tau}(X_{t+1} \mid \mathscr{F}_t) = \hat{\mu}_{t+1} + \hat{\sigma}_{t+1} \tilde{q}_{\tau}(Z)$ (for the GARCH-EVT method) where the difference lies in how $q_{\tau}(Z)$ is estimated. By comparison, the UGH method works directly on the series $X_t$, without filtering, the estimate then being $\bar{q}_{\tau}(X_{t+1} \mid \mathscr{F}_t) = \hat{q}_{\tau}(X)$, where $\hat{q}_{\tau}(X)$ is obtained as in Section~\ref{S:UGHstep} with the $X_t$ in place of the $\hat{Z}_t$. In addition, because the estimation of the constant $\rho$ is known to be a difficult problem, we make the following adjustment to the GARCH-UGH method. We calculate another version of the GARCH-UGH estimate where the estimator $\hat{\rho}_{k_{\rho}}$ is replaced throughout by the constant $-1$. If this other version has a number of VaR violations closer to the expected number of violations (which is known and equal to $3000(1-\tau)$ where $\tau$ is the VaR level), we retain this version. The use of the constant $-1$ has been advocated in the literature before, see for example Section $4.5.1$ in~\cite{beirlant2004}.

\begin{figure}
\begin{subfigure}{0.99\textwidth}
\includegraphics[width=0.99\linewidth, height=8.0cm]{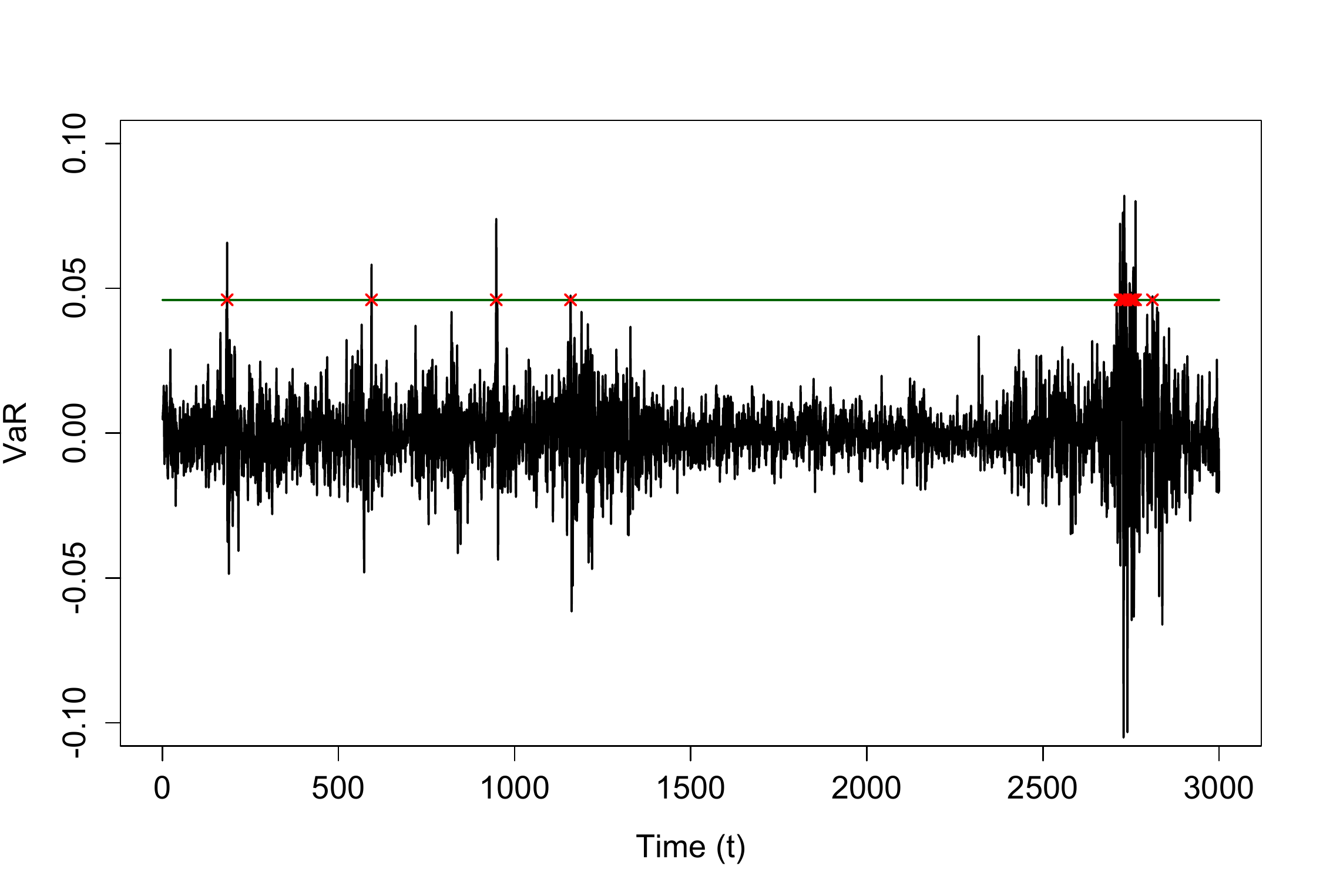} 
\caption{UGH approach }
\label{F:4.1}
\end{subfigure} \\
\begin{subfigure}{0.99\textwidth}
\includegraphics[width=0.99\linewidth, height=8.0cm]{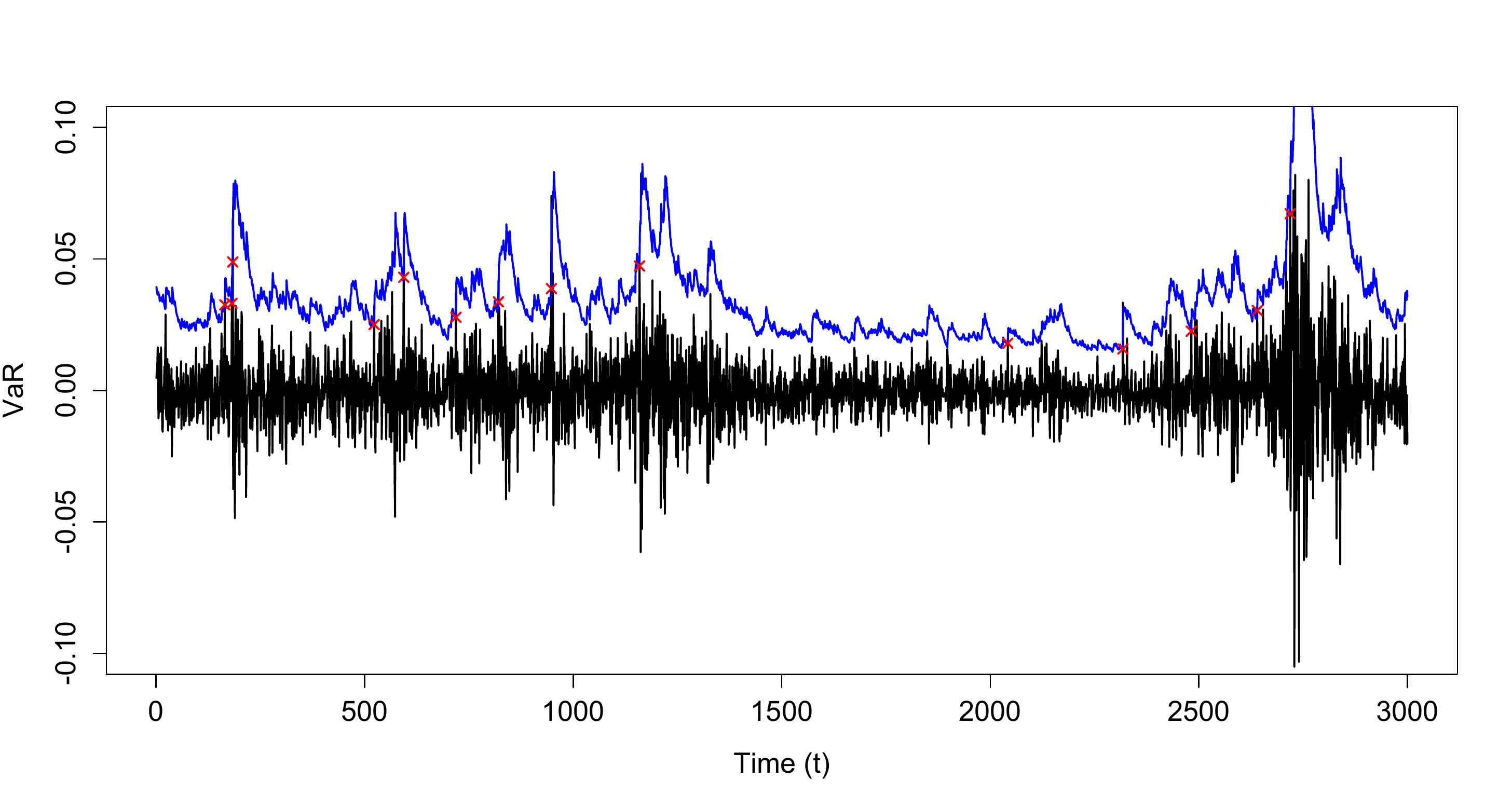}
\caption{GARCH-UGH approach}
\label{F:4.2}
\end{subfigure}

\caption{Twelve years (8 December 1997 to 9 November 2009) 
of in-sample backtesting of the DJ index, and $99.5\%$-VaR violations by (a) the UGH approach and (b) the GARCH-UGH approach when the top $15\%$ of observations are used for the estimation. Red cross marks denote the VaR violations.}
\label{F:4}
\end{figure}

Results from Tables \ref{T:2}-\ref{T:5} indicate that, on the basis of in-sample validation, the proposed GARCH-UGH approach is the most successful for estimating one-step ahead extreme VaRs that satisfy both unconditional and conditional coverage properties. Across all samples and in terms of number of VaR violations only, in $46$ out of $60$ cases our GARCH-UGH approach is closest to the mark, with the conventional GARCH-EVT performing worst overall. In addition, although the unfiltered UGH estimate is somewhat reasonable in terms of number of VaR violations, it is not appropriate because it lacks responsiveness to the time-varying volatility and volatility clustering: Figure \ref{F:4.1} illustrates that the non-dynamic nature of the UGH estimate leaves it unable to respond immediately to high volatility, and VaR violations tend to cluster. By contrast, the conditional VaR estimates obtained by our GARCH-UGH approach (Figure \ref{F:4.2}) clearly respond to the changing volatility with no clustering of VaR violations, while bias reduction results in closer numbers of VaR violations to the expected numbers than with the conventional GARCH-EVT. Numerically, the GARCH-UGH method never fails either the Kupiec or Christoffersen tests, whereas the GARCH-EVT method fails $7$ and $5$ times out of $60$ cases, respectively. The bias correction at the extreme value step appears to be very effective for the accurate estimation of one-step ahead dynamic extreme VaRs. It leads to results that seem less sensitive to the choice of sample fraction $k$ than the conventional GARCH-EVT method: see Tables \ref{T:2}-\ref{T:5}, where results appear to be consistently good across a large range of values of $k$.

\subsection{Out-of-sample dynamic extreme VaR estimation and backtesting}
\label{S:4.3}

We now focus on the out-of-sample estimation (that is, prediction) of one-step ahead VaR via the same three approaches, again at level $\tau \in \{0.99, 0.995, 0.999\}$. We consider the following samples of data: 
\begin{itemize}
\item The time period from 23 December 1993 to 9 November 2009 for the Dow Jones,  
\item The time period from 30 August 1993 to 16 July 2009 for the Nasdaq,
\item The time period from 14 May 1993 to 12 August 2009 for the Nikkei,
\item The time period from 2 January 2000 to 14 December 2010 for the JPY/GBP exchange rate.
\end{itemize}
In order to carry out this out-of-sample backtest, we adopt a rolling window estimation approach. Specifically, we first fix a testing window $W_T$ in each case, which corresponds to the periods of time considered in our in-sample evaluation (8 December 1997 to 9 November 2009 for the Dow Jones, 13 August 1997 to 16 July 2009 for the Nasdaq, 29 May 1997 to 12 August 2009 for the Nikkei, 28 September 2002 to 14 December 2010 for the JPY/GBP exchange rate). At each time $t$ in this testing window $W_T$, we use a window of length $W_E$ of prior information in order to predict the conditional VaR on time $t+1$ (with parameter estimates updated when the estimation window changes), which is then compared to the observed log-return on day $t+1$. Various choices of $W_E$ have been made in the literature: here we choose $W_E = 1000$ as in~\cite{mcneil2000}, corresponding to approximately four years of model calibration for each prediction with stock market data, and three years with exchange rate data.

\begin{figure}
\includegraphics[width=0.99\linewidth, height=6cm]{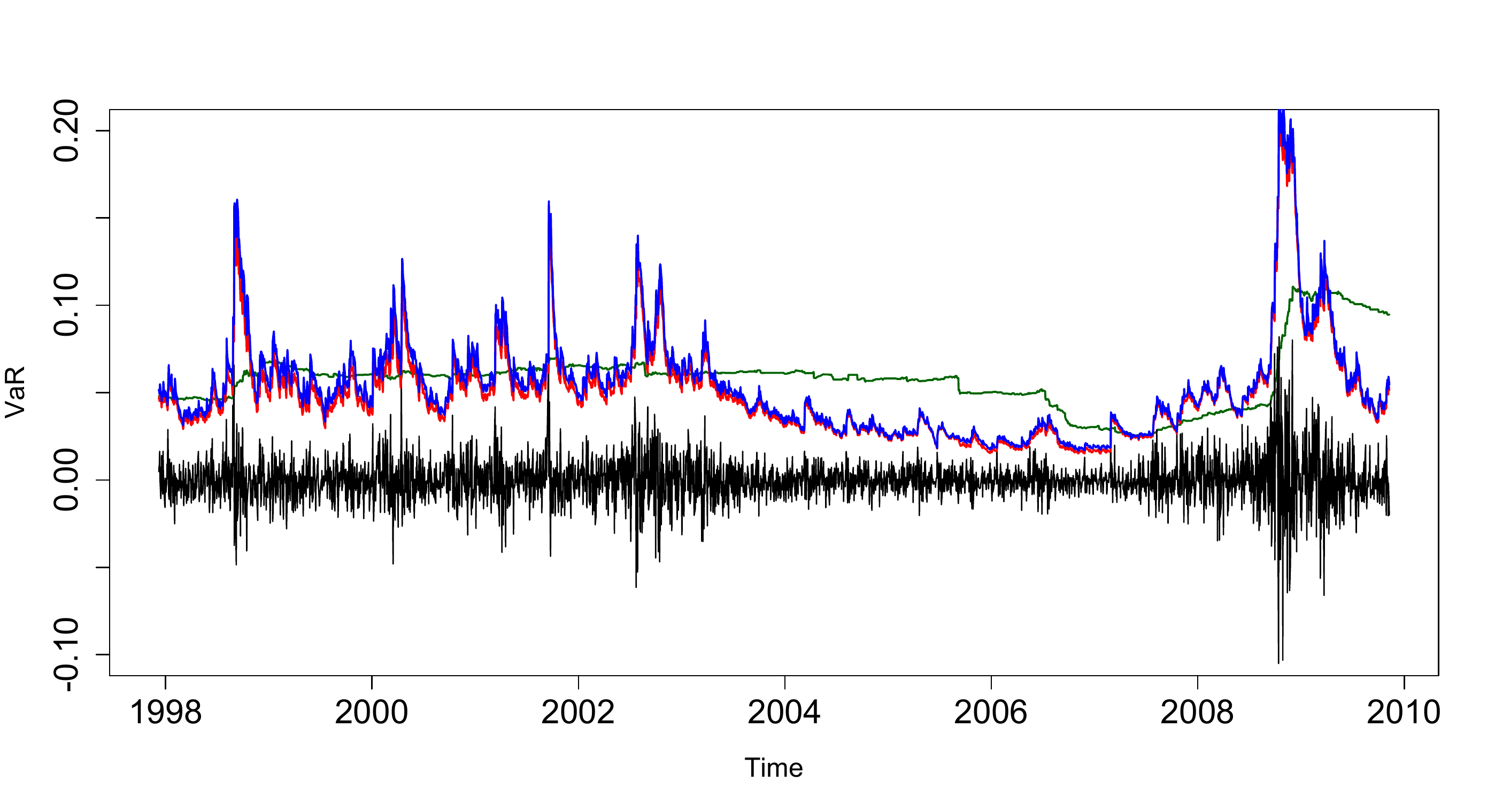} 
\caption{Out-of-sample backtesting of the DJ index from 8 December 1997 to 9 November 2009, 
and $99.9\%$-VaR estimates calculated using rolling estimation windows made of 1000 observations, 
with $k$ corresponding to the top $15\%$ observations from this window. GARCH-UGH (blue line), GARCH-EVT (red line) and UGH (dark green line) estimates are superimposed on the negative log-returns (black line).}
\label{F:5}
\end{figure}

\begin{figure}
\includegraphics[width=0.99\linewidth, height=6cm]{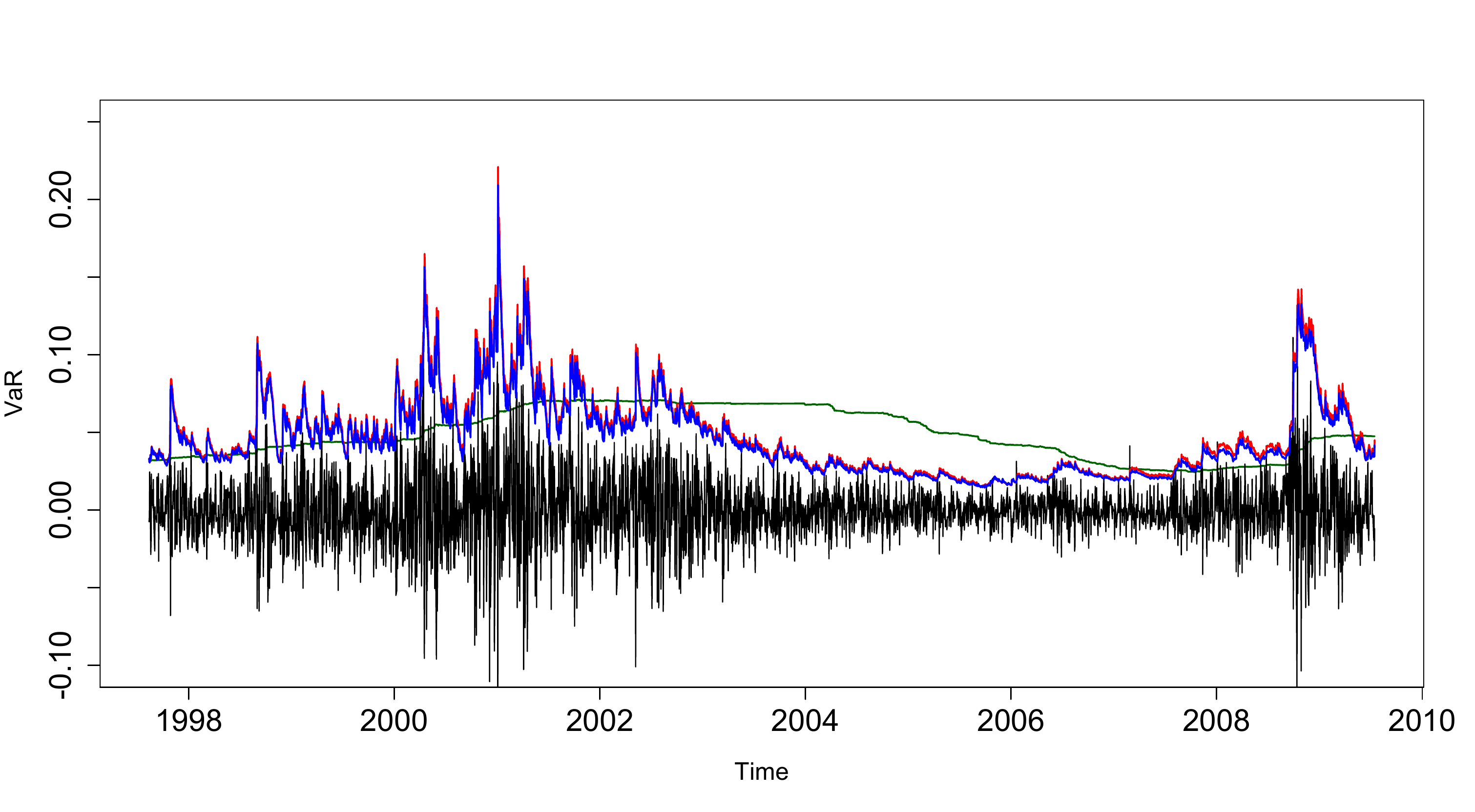} 
\caption{Out-of-sample backtesting of the NASDAQ index from 13 August 1997 to 16 July 2009, 
and $99.9\%$-VaR estimates calculated using rolling estimation windows made of 1000 observations, 
with $k$ corresponding to the top $20\%$ of observations from this window. GARCH-UGH (blue line), GARCH-EVT (red line) and UGH (dark green line) estimates are superimposed on the negative log-returns (black line).}
\label{F:6}
\end{figure}

\begin{figure}
\includegraphics[width=0.99\linewidth, height=6cm]{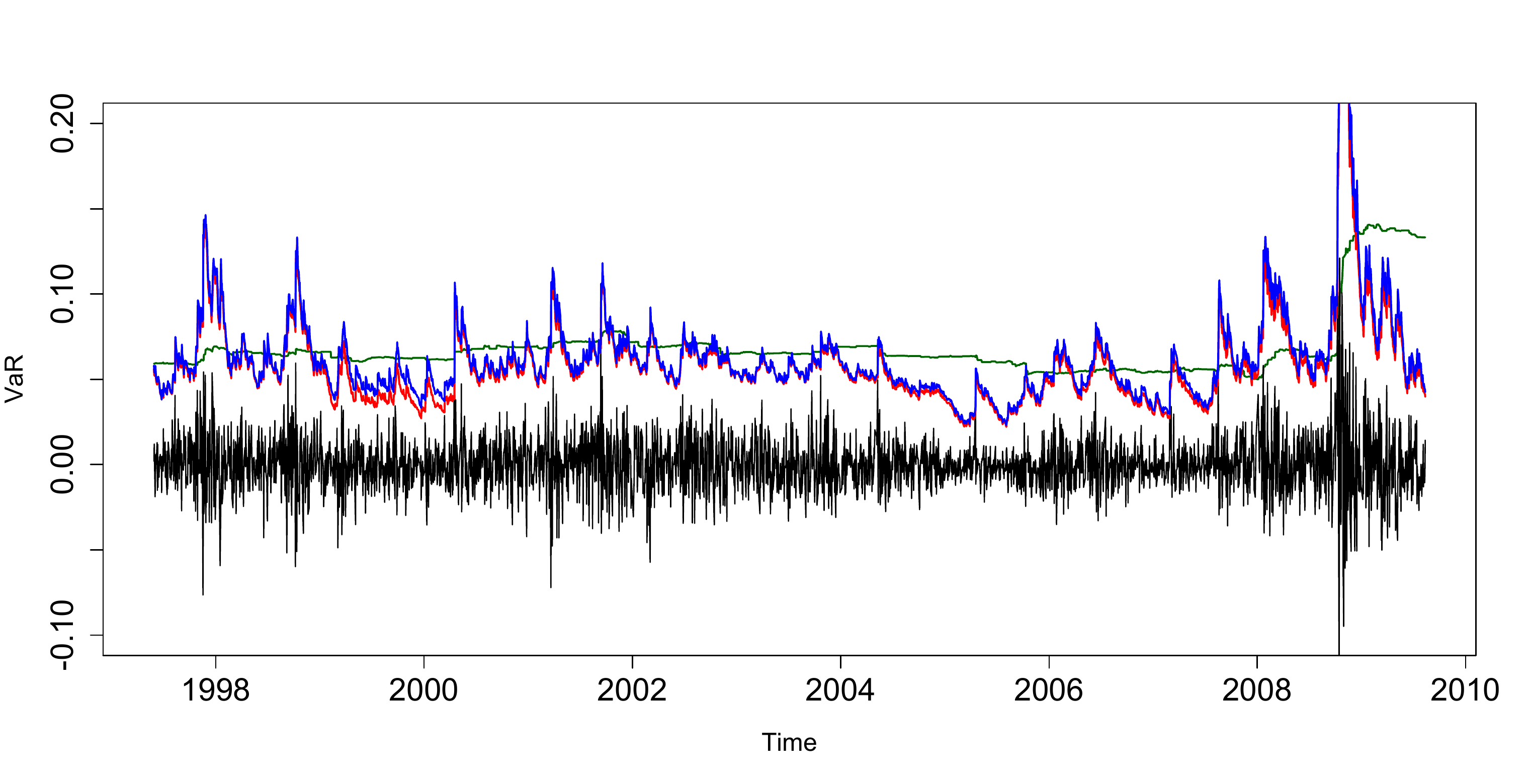} 
\caption{Out-of-sample backtesting of the NIKKEI index from 29 May 1997 to 12 August 2009, 
and $99.9\%$-VaR estimates calculated using rolling estimation windows made of 1000 observations, 
with $k$ corresponding to the top $10\%$ of observations from this window. GARCH-UGH (blue line), GARCH-EVT (red line) and UGH (dark green line) estimates are superimposed on the negative log-returns (black line).}
\label{F:7}
\end{figure}

\begin{figure}
\includegraphics[width=0.99\linewidth, height=6cm]{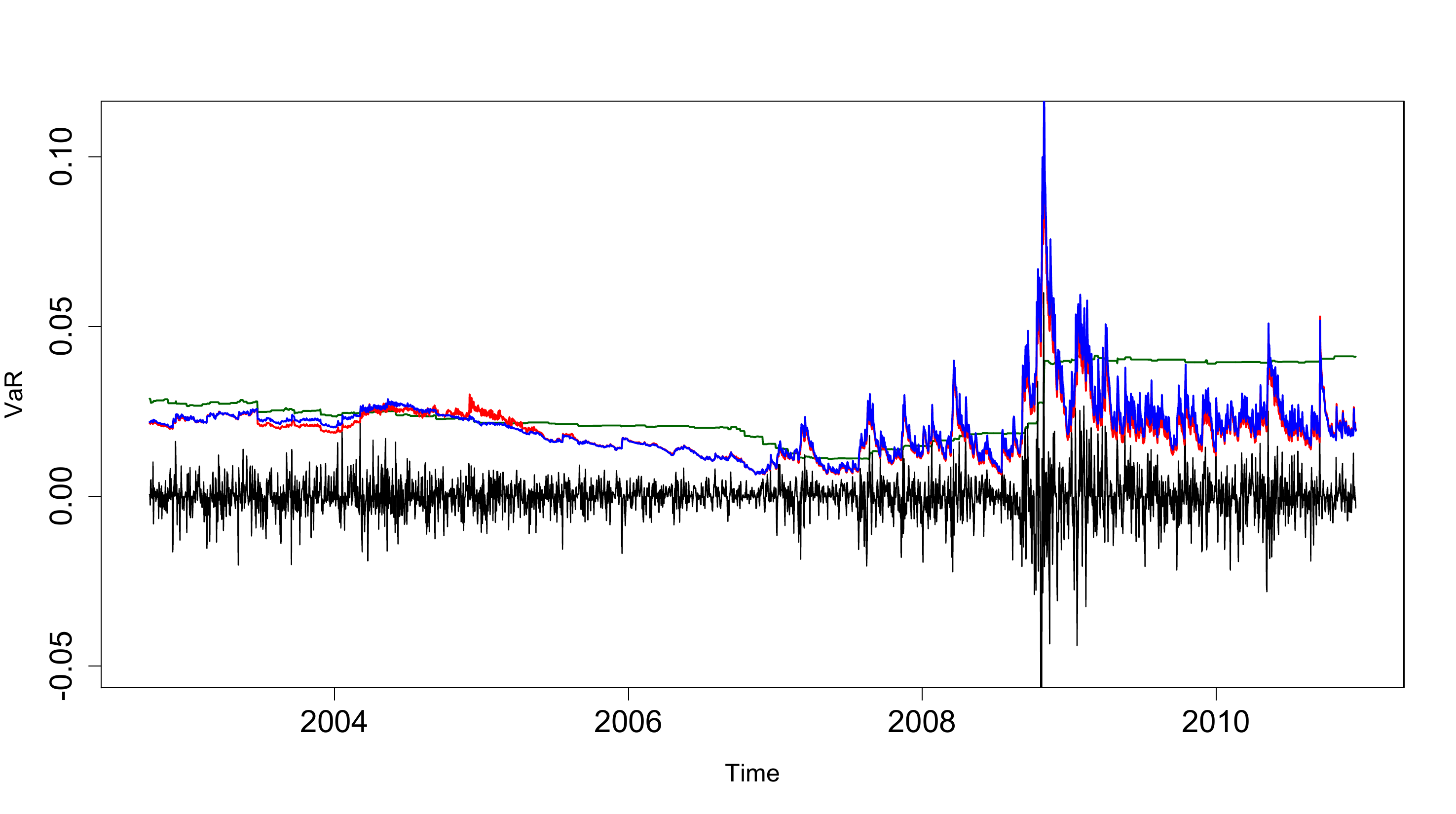} 
\caption{Out-of-sample backtesting of the JPY/GBP exchange rate from 28 September 2002 to 14 December 2010, 
and $99.9\%$-VaR estimates calculated using rolling estimation windows made of 1000 observations, 
with $k$ corresponding to the top $10\%$ of observations from this window. GARCH-UGH (blue line), GARCH-EVT (red line) and UGH (dark green line) estimates are superimposed on the negative log-returns (black line).}
\label{F:8}
\end{figure}

Tables \ref{T:6}-\ref{T:9} gather the numerical results. It can be seen that again, the suggested GARCH-UGH approach appears to be best overall. In $47$ out of $60$ cases, the GARCH-UGH approach yields the closest number of VaR violations to the theoretically expected numbers, while the unfiltered UGH method fares worst. Based on the Kupiec test, the GARCH-UGH approach fails twice, whereas the GARCH-EVT and UGH fail $6$ and $49$ times out of $60$ cases, respectively. On one occasion GARCH-UGH fails the Christoffersen test, while the GARCH-EVT and UGH methods fail $0$ and $43$ times out of $60$ cases. The GARCH-UGH typically performs better than other approaches except possibly when the top $5\%$ and $10\%$ of observations are used (for the choice of $k$); this is because the bias is not the dominating term in the bias-variance tradeoff when $k$ is small. The corresponding plots of out-of-sample backtesting are shown in Figures \ref{F:5}-\ref{F:8}, where it is clearly seen that the GARCH-UGH and GARCH-EVT estimates have the same dynamics, with the bias correction shifting the estimate upwards or downwards depending on the rolling estimation window.

\subsection{Conclusion from the empirical analysis}

We conclude from this empirical analysis that the proposed GARCH-UGH approach combining filtering via the AR(1)-GARCH(1,1) model and a semiparametric bias-reduced extreme value step provides better one-step ahead dynamic extreme VaR estimates for financial time series than the benchmark conventional GARCH-EVT approach of~\cite{mcneil2000}. This can be seen from both the in-sample and the out-of-sample estimations at several quantile levels $\tau$, including the very high $\tau=0.999$ corresponding to a $99.9\%$ VaR, and a large range of sample fractions $k$, due to the effect of the bias correction. In addition, from a computational standpoint, the GARCH-UGH method appears to be more reliable than the conventional GARCH-EVT approach, because the extreme value step is carried out using an automatic recipe for the estimation of the tail index and extreme quantile, whereas the GARCH-EVT approach relies on maximum likelihood estimation, which crucially depends on having a good starting value to fit a Generalized Pareto distribution. This also makes the GARCH-UGH method computationally cheaper than the GARCH-EVT competitor.

\section{Discussion}
\label{S:Discussion}

In this paper we introduce an extension of the two-step GARCH-EVT approach from \cite{mcneil2000} for extreme VaR estimation, based on a semiparametric bias-reduced extreme quantile estimator from~\cite{chavez2018,dehaan2016}. This differs from the other papers published in the econometric literature by introducing a finite-sample improvement at the extreme value step, rather than using a more complicated filter than the AR(1)-GARCH(1,1) filter.

Even though the Basel Committee on Banking Supervision recommends the use of VaR at high levels (see for example~\cite{basel2013}), the VaR itself has been criticized several times in the financial literature for two main reasons. First, the VaR only measures the frequency of observations below or above the predictor and not their magnitude: this means that, while it is known that $100(1-\tau)\%$ of losses will be higher than the VaR $q_{\tau}$ at level $\tau$, the VaR alone cannot give any further information about the size of these large losses. Second, the VaR is not a coherent risk measure in the sense of~\cite{artzner1999}, because it is not sub-additive in general, meaning that it does not abide by the intuitive diversification principle stating that a portfolio built on several financial assets carries less risk than a portfolio solely consisting of one of these assets. These two weaknesses pushed the Basel Committee to also recommend calculating the Expected Shortfall (or Conditional Value-at-Risk) as a complement or alternative to the VaR. In practice, this is hampered by the fact that the Expected Shortfall is not elicitable, and therefore the development of a simple backtesting methodology for the Expected Shortfall is not clear. This is why we believe that the accurate estimation of extreme VaR is still worth pursuing.

We highlight three possible directions for further investigations. The first one is that one could replace the AR(1)-GARCH(1,1) filter by a more sophisticated filter. Which filter should be used is not obvious: one could think about replacing the AR(1) part by an ARMA($p,q$) part, or the GARCH(1,1) part by a GARCH($p,q$) part (or a more complicated asymmetric version), or both. This may make it possible to even better account or the volatility dynamics, whose accurate estimation and prediction are key. The second one is the extension of our GARCH-UGH approach to the estimation of the multiple-step ahead conditional VaR. This is important, because certain regulations such as those advocated by \cite{basel2009} require the estimation of the 10-day ahead VaR at the $99\%$ confidence level, rather than merely the one-step ahead VaR. This is a challenging problem:~\cite{mcneil2000} tackle this question using a bootstrap methodology, but bootstrapping with heavy tails is known to be very difficult to calibrate. The development of an adaptation of the GARCH-UGH method to the multiple-step ahead setup which stays computationally manageable is well beyond the scope of the current paper. The third and final perspective is the estimation of alternative dynamic risk measures as a way of solving the two drawbacks of VaR that we highlighted in the previous paragraph. One candidate will be the expectile risk measure (see~\cite{newpow1987} for the original definition of expectiles in a regression context), which takes into account both the frequency of extreme observations and their magnitude, and is also shown to be a coherent and elicitable risk measure in~\cite{ziegel2016}. The use of expectiles has recently received substantial attention from the perspective of risk management as an alternative tool for quantifying tail risk (see for example~\cite{daouia2018,daogirstu2020}), but the case of dynamic estimation of extreme expectiles in a financial time series context has not been considered yet. Of course, there exists no universally preferred risk measure: the expectile only has an implicit formulation in general, and is more difficult to interpret than the VaR. The development of a GARCH-UGH-based method for the estimation of dynamic extreme expectiles will thus be an interesting complement to the present paper.

\section*{Acknowledgments}

The first author gratefully acknowledges the financial support from the SOKENDAI (Graduate University for Advanced Studies) under SOKENDAI Student Dispatch Program grant, and the Research Fellowships of Japan Society for the Promotion of Science for Young Scientists (Project number: 20J15188). Y. Kawasaki is supported by JSPS Grant-in-Aid for Scientific Research (18H00836, 19K01597, 20H01502) and by the ISM Cooperative Research Program 2020-ISMCRP-2038. G. Stupfler is supported by the French National Research Agency under the grant ANR-19-CE40-0013/ExtremReg project, and he also acknowledges support from an AXA Research Fund Award on ``Mitigating risk in the wake of the COVID-19 pandemic''. 


\bibliographystyle{plain} 
\bibliography{KSK_GARCH-UGH_final}

\begin{table}[htp]
\begin{center}
\caption{Summary of descriptive statistics and basic statistical tests for daily negative log-returns on DJ, NASDAQ, NIKKEI and JPY/GBP.}
\label{T:1}
\begin{tabular}{lclclclclcl}
\hline
 & DJ  & NASDAQ  & NIKKEI  & JPY/GBP  \\ \hline
Sample size & $4000$  & $4000$  & $4000$ & $4000$  \\ 
Mean & $-0.000250$  & $-0.000355$  & $-0.000169$  & $-0.0000557$  \\ 
Median & $-0.000460$  & $-0.00123$ & $-0.0000177$  & $0$  \\ 
Maximum & $0.0820$  & $0.111$  & $0.121$  & $0.0600$  \\
Minimum & $-0.105$  & $-0.172$  & $-0.132$  & $ -0.0640$  \\ 
Standard deviation & $0.0119$  & $0.0203$  & $0.0155$  & $0.00626$ \\ 
Skewness & $0.117$  & $-0.110$  & $0.175$  & $-0.586$  \\ 
Kurtosis & $8.096$  & $4.469$  & $5.579$  & $10.931$  \\ 
J-B test & $10933  \ast$  & $3337.3  \ast$  & $5207.9  \ast$  & $2014.6  \ast$  \\ 
& $ (0.0000)$  & $(0.0000)$  & $ (0.0000)$  & $ (0.0000)$  \\ 
$Q(1)$ & $13.159 \ast$   & $12.098  \ast$  & $6.680  \ast$  & $128.68  \ast$  \\ 
 & $ (0.000)$  & $ (0.001)$  & $ (0.010)$  & $ (0.000)$  \\ 
$Q(5)$ & $37.723  \ast$  & $37.62  \ast$ & $14.429  \ast$  & $146.47  \ast$  \\ 
& $(0.000)$  & $ (0.000)$ & $ (0.013)$  & $(0.000)$  \\ 
$Q(10)$ & $50.388  \ast$  & $42.192  \ast$  &  $23.023  \ast$  & $150.37  \ast$  \\ 
 & $ (0.000)$  & $ (0.000)$  &  $ (0.011)$  & $ (0.000)$  \\ 
$Q^{2}(1)$ & $131.54  \ast$  & $207.21  \ast$  & $248.32  \ast$  & $275.36  \ast$  \\
 & $ (0.000)$  & $ (0.000)$  & $ (0.000)$  & $ (0.000)$  \\
$Q^{2}(10)$ & $2613.3  \ast$  & $1907.5  \ast$  & $3183.4  \ast$  & $1650.4  \ast$  \\ 
 & $ (0.000)$  & $ (0.000)$  & $ (0.000)$  & $ (0.000)$  \\   
ADF test & $-15.782 {\ast\ast}$  & $-14.794 {\ast\ast}$  & $-15.967 {\ast\ast}$  & $-16.415 {\ast\ast}$  \\ \hline
\end{tabular}
\end{center}
\caption*{\footnotesize{Notes: A kurtosis greater than $3$ indicates that the dataset has heavier tails than a normal distribution. J-B stands for the Jarque-Bera test, $Q(n)$ and $Q^{2}(n)$ are the Ljung-Box tests for autocorrelation at lags $n$ in the negative log-return series and squared negative log-returns, respectively. The ADF test is the augmented Dickey-Fuller stationarity test statistic without trend. 
The $p$-values are given between brackets. ${\ast\ast}$, $\ast$ denote significance at $1\%$ and $5\%$ levels, respectively.}}
\end{table}

\begin{table}[htp]
\begin{center}
\caption{In-sample evaluations of one-step ahead conditional VaR estimates from 8 December 1997 to 9 November 2009 
at different quantile levels for the negative log-returns of DJ index by means of the number of VaR violations, unconditional and conditional coverage tests.}
\label{T:2}
\scalebox{0.75}{
\begin{tabular}{l c c c c c r}
\hline
 Testing window & 3000  &  &  &  &    \\  \hline
 $\%$ of top obs. used & 5$\%$  &10$\%$  &15$\%$  &20$\%$&  25$\%$ \\ \hline
DJ:&  &  &  &  &    \\
 \textit{0.999 Quantile}&  &  &  &  &    \\
 Expected&3  &3  &3  &3  &3    \\
 UGH& \textbf{4} &5 &\textbf{2} &\textbf{2}&\textbf{3} \\
        & (0.583, 0.855)& (0.292, 0.569)& (0.538, 0.826)& (0.538, 0.826)& (1.000, 0.997) \\
 GARCH-UGH & \textbf{2}&\textbf{2}&\textbf{2}&\textbf{2}& 2 \\
         & (0.538, 0.826)& (0.538, 0.826)& (0.538, 0.826)& (0.538, 0.826)& (0.538, 0.826) \\
 GARCH-EVT & \textbf{2}&\textbf{2}&\textbf{2}&\textbf{2}& 4    \\
        & (0.538, 0.826)& (0.538, 0.826)& (0.538, 0.826)& (0.538, 0.826)& (0.583, 0.855) \\
  \textit{0.995 Quantile}&  &  &  &  &    \\
 Expected&15  &15  &15  &15  &15    \\
 UGH& 18&18&\textbf{16}&18&20  \\
                & (0.452, 0.012)& (0.452, 0.012)& (0.798, 0.009)& (0.452, 0.012)& (0.218, 0.011) \\
 GARCH-UGH & \textbf{15}&\textbf{14}&\textbf{14}&\textbf{15}&\textbf{15} \\
                     & (1.000, 0.927)& (0.793, 0.905)& (0.793, 0.905)& (1.000, 0.927)& (1.000, 0.927) \\
 GARCH-EVT& 13&13&13&13&13 \\
                      & (0.596, 0.821)& (0.596, 0.821)& (0.596, 0.821)& (0.596, 0.821)& (0.596, 0.821) \\
  \textit{0.99 Quantile}&  &  &  &  &    \\
 Expected&30&30&30&30&30 \\
 UGH&34&34&34&36&39   \\
                       & (0.472, 0.018)& (0.472, 0.018)& (0.472, 0.018)& (0.286, 0.018)& (0.114, 0.014) \\
 GARCH-UGH &\textbf{27}&\textbf{28}&\textbf{29}&\textbf{31}&\textbf{33}  \\
                       & (0.576, 0.669)& (0.711, 0.717)& (0.854, 0.741)& (0.855, 0.711)& (0.588, 0.598) \\
 GARCH-EVT& 23&23&22&22&20 \\
                        & (0.180, 0.341)& (0.180, 0.341)& (0.123, 0.259)& (0.123, 0.259)& (0.050, 0.130) \\ \hline
\end{tabular}}
\end{center}
\caption*{\footnotesize{Notes: The closest numbers of VaR violations to theoretically expected ones are highlighted in bold. The p-values for the unconditional coverage test by \cite{kupiec1995} and conditional coverage test by \cite{christoffersen1998} at the 5$\%$ significance level are given in brackets in order.}}
\end{table}

\begin{table}[htp]
\begin{center}
\caption{In-sample evaluations of one-step ahead conditional VaR estimates from 13 August 1997 to 16 July 2009 
at different quantile levels for the negative log-returns of NASDAQ index by means of the number of VaR violations, unconditional and conditional coverage tests.}
\label{T:3}
\scalebox{0.75}{
\begin{tabular}{l c c c c c r}
\hline
 Testing window & 3000  &  &  &  &    \\  \hline
 $\%$ of top obs. used & 5$\%$  &10$\%$  &15$\%$  &20$\%$&  25$\%$ \\ \hline
NASDAQ:&  &  &  &  &    \\
 \textit{0.999 Quantile}&  &  &  &  &    \\
 Expected&3  &3  &3  &3  &3    \\
 UGH& \textbf{3} &1 &1 &1&1 \\
        & (1.000, 0.997)& (0.179, 0.406)& (0.179, 0.406)& (0.179, 0.406)& (0.179, 0.406) \\
 GARCH-UGH & 4&\textbf{4}&\textbf{4}&\textbf{4}& \textbf{2} \\
         & (0.583, 0.855)& (0.583, 0.855)& (0.583, 0.855)& (0.583, 0.855)& (0.538, 0.826) \\
 GARCH-EVT & 4&\textbf{4}&\textbf{4}&\textbf{4}& \textbf{4}    \\
        & (0.583, 0.855)& (0.583, 0.855)& (0.583, 0.855)& (0.583, 0.855)& (0.583, 0.855) \\
  \textit{0.995 Quantile}&  &  &  &  &    \\
 Expected&15  &15  &15  &15  &15    \\
 UGH& 21&21&21&19&21  \\
                & (0.143, 0.295)& (0.143, 0.295)& (0.143, 0.295)& (0.320, 0.541)& (0.143, 0.295) \\
 GARCH-UGH & \textbf{14}&\textbf{14}&\textbf{14}&\textbf{14}&\textbf{13} \\
                     & (0.793, 0.905)& (0.793, 0.905)& (0.793, 0.905)& (0.793, 0.905)& (0.596, 0.821) \\
 GARCH-EVT& 13&13&10&10&10 \\
                      & (0.596, 0.821)& (0.596, 0.821)& (0.168, 0.374)& (0.168, 0.374)& (0.168, 0.374) \\
  \textit{0.99 Quantile}&  &  &  &  &    \\
 Expected&30&30&30&30&30 \\
 UGH&\textbf{32}&\textbf{33}&\textbf{33}&\textbf{35}&37   \\
                       & (0.717, 0.609)& (0.588, 0.135)& (0.588, 0.135)& (0.371, 0.127)& (0.215, 0.106) \\
 GARCH-UGH &23&23&23&\textbf{25}&\textbf{25}  \\
                       & (0.180, 0.341)& (0.180, 0.341)& (0.180, 0.341)& (0.345, 0.519)& (0.345, 0.519) \\
 GARCH-EVT& 22&17&16&16&16 \\
                        & (0.123, 0.259)& (0.009, 0.031)& (0.005, 0.017)& (0.005, 0.017)& (0.005, 0.017) \\ \hline
\end{tabular}}
\end{center}
\caption*{\footnotesize{Notes: The closest numbers of VaR violations to theoretically expected ones are highlighted in bold. The p-values for the unconditional coverage test by \cite{kupiec1995} and conditional coverage test by \cite{christoffersen1998} at the 5$\%$ significance level are given in brackets in order.}}
\end{table}

\begin{table}[htp]
\begin{center}
\caption{In-sample evaluations of one-step ahead conditional VaR estimates from 29 May 1997 to 12 August 2009 
at different quantile levels for the negative log-returns of NIKKEI index by means of the number of VaR violations, unconditional and conditional coverage tests.}
\label{T:4}
\scalebox{0.75}{
\begin{tabular}{l c c c c c r}
\hline
 Testing window & 3000  &  &  &  &    \\  \hline
 $\%$ of top obs. used & 5$\%$  &10$\%$  &15$\%$  &20$\%$&  25$\%$ \\ \hline
NIKKEI:&  &  &  &  &    \\
 \textit{0.999 Quantile}&  &  &  &  &    \\
 Expected&3  &3  &3  &3  &3    \\
 UGH& \textbf{4} &\textbf{4} &\textbf{4} &\textbf{4}&\textbf{1} \\
        & (0.583, 0.855)& (0.583, 0.855)& (0.583, 0.855)&  (0.583, 0.855)& (0.179, 0.406) \\
 GARCH-UGH & \textbf{4}&\textbf{2}&\textbf{4}&\textbf{4}& \textbf{1} \\
         & (0.583, 0.885)& (0.538, 0.826)& (0.583, 0.885)& (0.583, 0.885)& (0.179, 0.406) \\
 GARCH-EVT & 5&5&5&5&\textbf{5}    \\
        & (0.292, 0.569)& (0.292, 0.569)& (0.292, 0.569)&  (0.292, 0.569)&  (0.292, 0.569) \\
  \textit{0.995 Quantile}&  &  &  &  &    \\
 Expected&15  &15  &15  &15  &15    \\
 UGH& \textbf{15}&\textbf{15}&\textbf{17}&18&21  \\
                & (1.000, 0.178)& (1.000, 0.178)& (0.612, 0.199)& (0.452, 0.190)& (0.143, 0.114) \\
 GARCH-UGH & 13&13&\textbf{13}&\textbf{13}&\textbf{12} \\
                     & (0.596, 0.821)& (0.596, 0.821)& (0.596, 0.821)& (0.596, 0.821)& (0.421, 0.689) \\
 GARCH-EVT& 13&12&12&12&\textbf{12} \\
                      & (0.596, 0.821)& (0.421, 0.689)& (0.421, 0.689)& (0.421, 0.689)& (0.421, 0.689) \\
  \textit{0.99 Quantile}&  &  &  &  &    \\
 Expected&30&30&30&30&30 \\
 UGH&\textbf{32}&\textbf{32}&\textbf{34}&36&38   \\
                       & (0.717, 0.609)& (0.717 0.609)& (0.472, 0.562)& (0.286 0.427)& (0.159, 0.297) \\
 GARCH-UGH &26&25&\textbf{26}&\textbf{31}&\textbf{28} \\
                       & (0.453, 0.601)& (0.345, 0.519)& (0.453, 0.601)& (0.855, 0.711)& (0.711, 0.666) \\
 GARCH-EVT& 25&24&21&19&18 \\
                        & (0.345, 0.287)& (0.254, 0.430)& (0.081, 0.188)& (0.030, 0.085)& (0.017, 0.049) \\ \hline
\end{tabular}}
\end{center}
\caption*{\footnotesize{Notes: The closest numbers of VaR violations to theoretically expected ones are highlighted in bold. The p-values for the unconditional coverage test by \cite{kupiec1995} and conditional coverage test by \cite{christoffersen1998} at the 5$\%$ significance level are given in brackets in order.}}
\end{table}

\begin{table}[htp]
\begin{center}
\caption{In-sample evaluations of one-step ahead conditional VaR estimates from 28 September 2002 to 14 December 2010 
at different quantile levels for the negative log-returns of JPY/GBP exchange rate by means of the number of VaR violations, unconditional and conditional coverage tests.}
\label{T:5}
\scalebox{0.75}{
\begin{tabular}{l c c c c c r}
\hline
 Testing window & 3000  &  &  &  &    \\ \hline
 $\%$ of top obs. used & 5$\%$  &10$\%$  &15$\%$  &20$\%$&  25$\%$ \\ \hline
JPY/GBP:&  &  &  &  &    \\
 \textit{0.999 Quantile}&  &  &  &  &    \\
 Expected&3  &3  &3  &3  &3    \\
 UGH& 2 &2 &1 &1&1 \\
        & (0.538, 0.826)& (0.538, 0.826)& (0.179, 0.406)& (0.179, 0.406)& (0.179, 0.406) \\
 GARCH-UGH & \textbf{3}&2&\textbf{3}&2& 2 \\
         & (1.000, 0.997)& (0.538, 0.826)& (1.000, 0.997)& (0.538, 0.826)& (0.538, 0.826) \\
 GARCH-EVT & \textbf{3}&\textbf{3}&\textbf{3}&\textbf{3}& \textbf{3}    \\
        & (1.000, 0.997)& (1.000, 0.997)& (1.000, 0.997)& (1.000, 0.997)& (1.000, 0.997) \\
  \textit{0.995 Quantile}&  &  &  &  &    \\
 Expected&15  &15  &15  &15  &15    \\
 UGH& \textbf{16}&17&\textbf{16}&18&28  \\
                & (0.798, 0.195)& (0.612, 0.199)& (0.798 0.195)& (0.452, 0.190)& (0.003, 0.000) \\
 GARCH-UGH & \textbf{16}&\textbf{14}&\textbf{14}&\textbf{14}&\textbf{16} \\
                     & (0.798, 0.888)& (0.793, 0.905)& (0.793, 0.905)& (0.793, 0.905)& (0.798, 0.888) \\
 GARCH-EVT& 11&11&11&11&10 \\
                      & (0.277, 0.532)& (0.277, 0.532)& (0.277, 0.532)& (0.277, 0.532)& (0.168, 0.374) \\
  \textit{0.99 Quantile}&  &  &  &  &    \\
 Expected&30&30&30&30&30 \\
 UGH&38&40&41&41&46   \\
                       & (0.159, 0.002)& (0.081, 0.002)& (0.056, 0.001)& (0.056, 0.001)& (0.006, 0.001) \\
 GARCH-UGH &\textbf{31}&32&\textbf{31}&\textbf{29}&\textbf{22}  \\
                       & (0.855, 0.612)& (0.717, 0.609)& (0.855, 0.612)& (0.854, 0.556)& (0.123, 0.259) \\
 GARCH-EVT& \textbf{29}&\textbf{29}&28&24&\textbf{22} \\
                        & (0.854, 0.556)& (0.854, 0.556)& (0.711, 0.501)& (0.254, 0.430)& (0.123, 0.259) \\ \hline
\end{tabular}}
\end{center}
\caption*{\footnotesize{Notes: The closest numbers of VaR violations to theoretically expected ones are highlighted in bold. The p-values for the unconditional coverage test by \cite{kupiec1995} and conditional coverage test by \cite{christoffersen1998} at the 5$\%$ significance level are given in brackets in order.}}
\end{table}

\begin{table}[htp]
\begin{center}
\caption{Out-of-sample evaluations of one-step ahead conditional VaR estimates from 8 December 1997 to 9 November 2009 
at different quantile levels for the negative log-returns of DJ index by means of the number of VaR violations, unconditional and conditional coverage tests.}
\label{T:6}
\scalebox{0.75}{
\begin{tabular}{l c c c c c r}
\hline
 Testing window & 3000  &  &  &  &    \\ 
 Estimation window & 1000  &  &  &  &    \\ \hline
 $\%$ of top obs. used & 5$\%$  &10$\%$  &15$\%$  &20$\%$&  25$\%$ \\ \hline
DJ:&  &  &  &  &    \\
 \textit{0.999 Quantile}&  &  &  &  &    \\
 Expected&3  &3  &3  &3  &3    \\
 UGH& 10 &9 &9 &7 &6 \\
        & (0.001, 0.006)& (0.005, 0.020)& (0.005, 0.020)& (0.049, 0.142)& (0.128, 0.310) \\
 GARCH-UGH & \textbf{3}&\textbf{3}&\textbf{3}&\textbf{3}& \textbf{3} \\
         & (1.000, 0.997)& (1.000, 0.997)& (1.000, 0.997)& (1.000, 0.997)& (1.000, 0.997) \\
 GARCH-EVT & \textbf{3}&4&4&4& 4    \\
        & (1.000, 0.997)& (0.583, 0.885)& (0.583, 0.885)& (0.583, 0.885)& (0.583, 0.855) \\
  \textit{0.995 Quantile}&  &  &  &  &    \\
 Expected&15  &15  &15  &15  &15    \\
 UGH& 40&40&40&36&29  \\
                & (0.000, 0.000)& (0.000, 0.000)& (0.000, 0.000)& (0.000, 0.000)& (0.001, 0.001) \\
 GARCH-UGH & \textbf{19}&\textbf{18}&\textbf{18}&\textbf{16}&\textbf{14} \\
                     & (0.320, 0.541)& (0.452, 0.676)& (0.452, 0.676)& (0.798, 0.888)& (0.793, 0.905) \\
 GARCH-EVT& \textbf{19}&\textbf{18}&\textbf{18}&17&17 \\
                      & (0.320, 0.541)& (0.452, 0.676)& (0.452, 0.676)& (0.612, 0.798)& (0.612, 0.798) \\
  \textit{0.99 Quantile}&  &  &  &  &    \\
 Expected&30&30&30&30&30 \\
 UGH&62&64&63&63&61   \\
                       & (0.000, 0.000)& (0.000, 0.000)& (0.000, 0.000)& (0.000, 0.000)& (0.000, 0.000) \\
 GARCH-UGH &\textbf{33}&35&32&\textbf{31}&\textbf{28}  \\
                       & (0.588, 0.598)& (0.371, 0.433)& (0.717, 0.663)& (0.855, 0.711)& (0.711, 0.717) \\
 GARCH-EVT& \textbf{33}&\textbf{30}&\textbf{30}&28&27 \\
                        & (0.588, 0.598)& (1.000, 0.738)& (1.000, 0.738)& (0.711, 0.717)& (0.576, 0.669) \\ \hline
\end{tabular}}
\end{center}
\caption*{\footnotesize{Notes: The closest numbers of VaR violations to theoretically expected ones are highlighted in bold. The p-values for the unconditional coverage test by \cite{kupiec1995} and conditional coverage test by \cite{christoffersen1998} at the 5$\%$ significance level are given in brackets in order.}}
\end{table}

\begin{table}[htp]
\begin{center}
\caption{Out-of-sample evaluations of one-step ahead conditional VaR estimates from 13 August 1997 to 16 July 2009 
at different quantile levels for the negative log-returns of NASDAQ index by means of the number of VaR violations, unconditional and conditional coverage tests.}
\label{T:7}
\scalebox{0.75}{
\begin{tabular}{l c c c c c r}
\hline
 Testing window & 3000  &  &  &  &    \\ 
 Estimation window & 1000  &  &  &  &    \\ \hline
 $\%$ of top obs. used & 5$\%$  &10$\%$  &15$\%$  &20$\%$&  25$\%$ \\ \hline
NASDAQ:&  &  &  &  &    \\
 \textit{0.999 Quantile}&  &  &  &  &    \\
 Expected&3  &3  &3  &3  &3    \\
 UGH& 10 &8&7 &\textbf{4}&\textbf{3} \\
        & (0.001, 0.006)& (0.017, 0.057)& (0.049, 0.142)& (0.583, 0.855)& (1.000, 0.997) \\
 GARCH-UGH & \textbf{6}&\textbf{5}&\textbf{5}&\textbf{4}& \textbf{3} \\
         & (0.128, 0.370)& (0.292, 0.569)& (0.292, 0.569)& (0.583, 0.855)& (1.000, 0.997) \\
 GARCH-EVT & 7&7&7&7& 7    \\
        & (0.049, 0.142)& (0.049, 0.142)& (0.049, 0.142)& (0.049, 0.142)& (0.049, 0.142) \\
  \textit{0.995 Quantile}&  &  &  &  &    \\
 Expected&15  &15  &15  &15  &15    \\
 UGH& 39&37&35&36&40  \\
                & (0.000, 0.000)& (0.000, 0.000)& (0.000, 0.000)& (0.000, 0.000)& (0.000, 0.000) \\
 GARCH-UGH & 20&17&\textbf{15}&\textbf{16}&\textbf{13} \\
                     & (0.218, 0.410)& (0.612, 0.798)& (1.000, 0.927)& (0.798, 0.888)& (0.596, 0.821) \\
 GARCH-EVT& \textbf{16}&\textbf{14}&13&13&\textbf{13} \\
                      & (0.798, 0.888)& (0.793, 0.905)& (0.596, 0.821)& (0.596, 0.821)& (0.596, 0.821) \\
  \textit{0.99 Quantile}&  &  &  &  &    \\
 Expected&30&30&30&30&30 \\
 UGH&74&74&70&65&62   \\
                       & (0.000, 0.000)& (0.000, 0.000)& (0.000, 0.000)& (0.000, 0.000)& (0.000, 0.000) \\
 GARCH-UGH &34&35&\textbf{31}&\textbf{30}&\textbf{25}  \\
                       & (0.427, 0.544)& (0.371, 0.490)& (0.855, 0.612)& (1.000, 0.594)& (0.345, 0.287) \\
 GARCH-EVT& \textbf{31}&\textbf{28}&28&24&23 \\
                        & (0.855, 0.612)& (0.711, 0.501)& (0.711, 0.501)& (0.254, 0.430)& (0.180, 0.341) \\ \hline
\end{tabular}}
\end{center}
\caption*{\footnotesize{Notes: The closest numbers of VaR violations to theoretically expected ones are highlighted in bold. The p-values for the unconditional coverage test by \cite{kupiec1995} and conditional coverage test by \cite{christoffersen1998} at the 5$\%$ significance level are given in brackets in order.}}
\end{table}

\begin{table}[htp]
\begin{center}
\caption{Out-of-sample evaluations of one-step ahead conditional VaR estimates from 29 May 1997 to 12 August 2009 
at different quantile levels for the negative log-returns of NIKKEI index by means of the number of VaR violations, unconditional and conditional coverage tests.}
\label{T:8}
\scalebox{0.75}{
\begin{tabular}{l c c c c c r}
\hline
 Testing window & 3000  &  &  &  &    \\ 
 Estimation window & 1000  &  &  &  &    \\ \hline
 $\%$ of top obs. used & 5$\%$  &10$\%$  &15$\%$  &20$\%$&  25$\%$ \\ \hline
NIKKEI:&  &  &  &  &    \\
 \textit{0.999 Quantile}&  &  &  &  &    \\
 Expected&3  &3  &3  &3  &3    \\
 UGH& 7 &6 &6 &\textbf{5}&\textbf{5} \\
        & (0.049, 0.142)& (0.128, 0.310)& (0.128, 0.310)&  (0.292, 0.569)& (0.292, 0.569) \\
 GARCH-UGH & \textbf{4}&\textbf{3}&\textbf{2}&\textbf{2}& \textbf{1} \\
         & (0.583, 0.855)& (1.000, 0.997)& (0.538, 0.826)& (1.000, 0.997)& (0.179, 0.406) \\
 GARCH-EVT & 5&4&6&6&6    \\
        & (0.292, 0.569)& (0.583, 0.855)& (0.128, 0.310)&  (0.128, 0.310)&  (0.128, 0.310) \\
  \textit{0.995 Quantile}&  &  &  &  &    \\
 Expected&15  &15  &15  &15  &15    \\
 UGH& 34&34&34&30&23 \\
                & (0.000, 0.000)& (0.000, 0.000)& (0.000, 0.000)& (0.000, 0.000)& (0.055, 0.062) \\
 GARCH-UGH & \textbf{15}&\textbf{15}&\textbf{15}&\textbf{15}&\textbf{12} \\
                     & (1.000, 0.927)& (1.000, 0.927)& (1.000, 0.927)& (1.000, 0.927)& (0.421, 0.689) \\
 GARCH-EVT& 13&14&13&12&\textbf{12} \\
                      & (0.596, 0.821)& (0.793, 0.905)& (0.596, 0.821)& (0.421, 0.689)& (0.421, 0.689) \\
  \textit{0.99 Quantile}&  &  &  &  &    \\
 Expected&30&30&30&30&30 \\
 UGH&46&47&46&45&53   \\
                       & (0.006, 0.011)& (0.004, 0.007)& (0.004, 0.007)& (0.010, 0.015)& (0.000 0.000) \\
 GARCH-UGH &33&33&\textbf{33}&\textbf{30}&36 \\
                       & (0.588, 0.598)& (0.588, 0.598)& (0.588, 0.598)& (1.000, 0.738)& (0.286, 0.365) \\
 GARCH-EVT& \textbf{32}&\textbf{29}&\textbf{27}&27&\textbf{26} \\
                        & (0.717, 0.663)& (0.854, 0.741)& (0.576, 0.669)& (0.576, 0.669)& (0.453, 0.601) \\ \hline
\end{tabular}}
\end{center}
\caption*{\footnotesize{Notes: The closest numbers of VaR violations to theoretically expected ones are highlighted in bold. The p-values for the unconditional coverage test by \cite{kupiec1995} and conditional coverage test by \cite{christoffersen1998} at the 5$\%$ significance level are given in brackets in order.}}
\end{table}

\begin{table}[htp]
\begin{center}
\caption{Out-of-sample evaluations of one-step ahead conditional VaR estimates from 28 September 2002 to 14 December 2010 
at different quantile levels for the negative log-returns of JPY/GBP exchange rate by means of the number of VaR violations, unconditional and conditional coverage tests.}
\label{T:9}
\scalebox{0.75}{
\begin{tabular}{l c c c c c r}
\hline
 Testing window & 3000  &  &  &  &    \\ 
 Estimation window & 1000  &  &  &  &    \\ \hline
 $\%$ of top obs. used & 5$\%$  &10$\%$  &15$\%$  &20$\%$&  25$\%$ \\ \hline
JPY/GBP:&  &  &  &  &    \\
 \textit{0.999 Quantile}&  &  &  &  &    \\
 Expected&3  &3  &3  &3  &3    \\
 UGH& 7 &7 &6 &\textbf{4}&\textbf{4} \\
        & (0.049, 0.142)& (0.049, 0.142)& (0.128, 0.310)& (0.583, 0.855)& (0.583, 0.855) \\
 GARCH-UGH & \textbf{3}&\textbf{2}&\textbf{2}&\textbf{2}& \textbf{2} \\
         & (1.000, 0.997)& (0.538, 0.826)& (0.538, 0.826)& (0.538, 0.826)& (0.538, 0.826) \\
 GARCH-EVT & 6&5&5&6& 7    \\
        & (0.128, 0.310)& (0.292, 0.569)& (0.292, 0.569)& (0.128, 0.310)& (0.049, 0.142) \\
  \textit{0.995 Quantile}&  &  &  &  &    \\
 Expected&15  &15  &15  &15  &15    \\
 UGH& 25&27&27&34&45  \\
                & (0.018, 0.028)& (0.005, 0.010)& (0.005, 0.010)& (0.000, 0.000)& (0.000, 0.000) \\
 GARCH-UGH & 21&\textbf{18}&\textbf{15}&\textbf{14}&\textbf{12} \\
                     & (0.143, 0.295)& (0.452, 0.676)& (1.000, 0.927)& (0.793, 0.905)& (0.421, 0.689) \\
 GARCH-EVT& \textbf{19}&19&20&20&20 \\
                      & (0.320, 0.541)& (0.320, 0.541)& (0.219, 0.410)& (0.219, 0.410)& (0.219, 0.410) \\
  \textit{0.99 Quantile}&  &  &  &  &    \\
 Expected&30&30&30&30&30 \\
 UGH&47&56&55&59&67   \\
                       & (0.004, 0.000)& (0.000, 0.000)& (0.000, 0.000)& (0.000, 0.000)& (0.000, 0.000) \\
 GARCH-UGH &42&46&40&\textbf{38}&\textbf{34}  \\
                       & (0.038, 0.064)& (0.006, 0.012)& (0.081, 0.127)& (0.159, 0.227)& (0.472, 0.523) \\
 GARCH-EVT& \textbf{38}&\textbf{37}&\textbf{38}&\textbf{38}&36 \\
                        & (0.159, 0.227)& (0.215, 0.292)& (0.159, 0.227)& (0.159, 0.227)& (0.286, 0.365) \\ \hline
\end{tabular}}
\end{center}
\caption*{\footnotesize{Notes: The closest numbers of VaR violations to theoretically expected ones are highlighted in bold. The p-values for the unconditional coverage test by \cite{kupiec1995} and conditional coverage test by \cite{christoffersen1998} at the 5$\%$ significance level are given in brackets in order.}}
\end{table}

\end{document}